\documentclass[12pt]{article}
\pdfoutput=1
\usepackage[utf8]{inputenc}

\usepackage{xcolor}
\usepackage{amsmath,amssymb,mathrsfs,bm,setspace,xspace,soul,empheq}  
\usepackage{graphicx}
\usepackage{physics}
\usepackage{colonequals} 
\usepackage{float}  
\usepackage{comment}    
\usepackage{siunitx}       
\usepackage[tight]{subfigure}
\usepackage{tablefootnote}
\usepackage{dsfont}
\usepackage{amsthm}

\usepackage[font={small}]{caption}

\usepackage{braket}
\usepackage{color}  
\usepackage{dcolumn}
\usepackage{multirow}            
\usepackage{geometry}         
\usepackage{tabularx}
\definecolor{darkblue}{rgb}{0.1,0.1,.7}
\usepackage[colorlinks, linkcolor=darkblue, citecolor=darkblue, urlcolor=darkblue, linktocpage,backref=page]{hyperref} 
\renewcommand*{\backref}[1]{}
\renewcommand*{\backrefalt}[4]{%
	\ifcase #1 (Not cited.)%
	\or        (Cited on p.~#2.)%
	\else      (Cited on pp.~#2.)%
	\fi}

\usepackage{amsmath,amssymb,graphicx,enumerate,bbm}
\usepackage{booktabs}
\usepackage{geometry}
\usepackage{tikz}
\usetikzlibrary{spy}

\definecolor{martenorange}{rgb}{0.91, 0.41, 0.17}

\usepackage[english]{babel}
\usepackage{csquotes}
\MakeOuterQuote{"}

\makeatletter
\newcommand{\subalign}[1]{%
	\vcenter{%
		\Let@ \restore@math@cr \default@tag
		\baselineskip\fontdimen10 \scriptfont\tw@
		\advance\baselineskip\fontdimen12 \scriptfont\tw@
		\lineskip\thr@@\fontdimen8 \scriptfont\thr@@
		\lineskiplimit\lineskip
		\ialign{\hfil$\m@th\scriptstyle##$&$\m@th\scriptstyle{}##$\hfil\crcr
			#1\crcr
		}%
	}%
}
\makeatother
\usepackage{enumitem}

\def\eps{\epsilon}
\newcommand{\beq}{\begin{equation}} 
\newcommand{\eeq}{\end{equation}}
 
\def\nn{\nonumber}

\def\calN {{\cal N}}

\def\ge{\geqslant}
\def\le{\leqslant}

\def\leq{\leqslant}
\def\<{\langle}
\def\>{\rangle}

\newcommand{\myinclude}[2][]{\raisebox{1ex-0.5\height}{\includegraphics[#1]{#2}}}

\newtheorem{proposition}{Proposition}[section]
\newtheorem{lemma}{Lemma}[section]
\newtheorem{theorem}{Theorem}[section]

\theoremstyle{remark}
\newtheorem{remark}{Remark}[section]
\newcommand {\no}{\noindent}
\renewcommand \qed{\vrule height5pt width5pt}

\numberwithin{equation}{section}
\usepackage{titlesec}
\titleformat*{\section}{\large\bfseries}
\titleformat*{\subsection}{\normalsize\bfseries}
\titleformat*{\subsubsection}{\normalsize\it}
\titleformat*{\paragraph}{\normalsize\bfseries}
\titleformat*{\subparagraph}{\normalsize\bfseries}

\usepackage{cancel}
\usepackage{soul}

\begin{document}
	\vspace*{-.6in} \thispagestyle{empty}
	\begin{flushright}
		
	\end{flushright}
	\vspace{1cm} 
	{\Large
		\begin{center}
			{\bf Tensor RG approach to high-temperature\\ fixed point}
		\end{center}
	}
	\vspace{1cm}
	\begin{center}
		{\bf Tom Kennedy$^a$,  Slava Rychkov$^{b,c}$
		}\\[2cm] 
		{
			$^a$ Department of Mathematics, University of Arizona,
			Tucson, AZ 85721, USA \\
			$^b$  Institut des Hautes \'Etudes Scientifiques,
			91440 Bures-sur-Yvette, France\\
			$^c$  
			Laboratoire de Physique de l'Ecole normale sup\'erieure, ENS,\\ 
			{\small Universit\'e PSL, CNRS, Sorbonne Universit\'e,
				Universit\'e de Paris,} F-75005 Paris, France
		}
		\vspace{1cm}
	\end{center}
	
	\vspace{4mm}
	\begin{abstract}
          We study a renormalization group (RG) map for tensor networks that
          include two-dimensional lattice spin systems such as the Ising
          model. Numerical studies of such RG maps have been quite successful
          at reproducing the known critical behavior.
          In those numerical studies the RG map must be truncated
          to keep the dimension of the legs of the tensors bounded.
          Our tensors act on an infinite-dimensional Hilbert space, and our
          RG map does not involve any truncations. Our RG map has a trivial
          fixed point which represents the high-temperature fixed point. We
          prove that if we start with a tensor that is close to this fixed point
          tensor, then the iterates of the RG map converge in the
          Hilbert-Schmidt norm to the fixed point tensor. 
          It is important to emphasize that this statement is not true for the
          simplest tensor network RG map in which one simply contracts
          four copies of the tensor to define the renormalized tensor. 
          The linearization of this simple RG map
          about the fixed point is not a contraction due to the presence of
          so-called CDL tensors. 
          Our work provides a first
          step towards the important problem of the rigorous study of RG maps
          for tensor networks in a neighborhood of the critical point.
	\end{abstract}
	
	\vspace{.2in}
	\vspace{.3in}
	\hspace{0.2cm} July 2021
	
	\newpage

	{\small
		\parskip=-0.1em
		\tableofcontents
	}
	
	
	\section{Introduction}
	
	In this paper we will study the tensor network approach to the renormalization group (RG) in the high-temperature phase of 2D lattice models. In the
	future, this approach could be used to rigorously describe continuous phase
	transitions in the Ising and other models (both in 2D and in 3D). Our work may
	be seen as a warmup towards that more interesting but much harder problem.
	
\begin{figure}[h]
		\centering
		\includegraphics[scale=0.5]{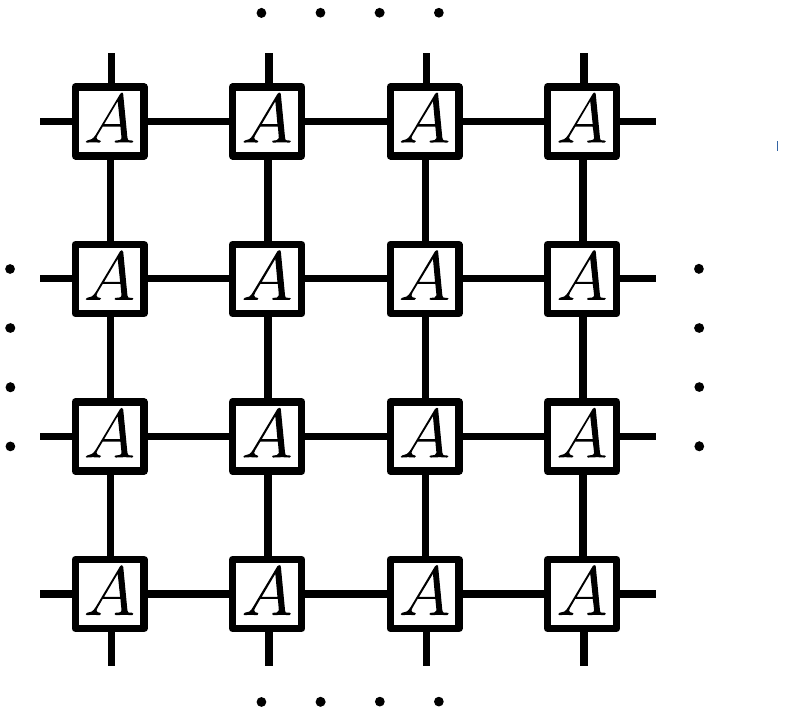}
		\caption{\label{fig-intro1}A tensor network built out of a four-tensor $A$. It is assumed that, after a large number of periodic repetitions, the outgoing bonds on network sides are joined pairwise, so that no uncontracted indices remain.}
	\end{figure}

	A tensor network is a periodic arrangement of tensors, with links indicating
	which tensor indices have to be contracted. See Fig.~\ref{fig-intro1} for a tensor network
	built out of a four-index tensor $A_{i j k l}$. We can naturally view such a
	tensor as a multilinear map:
	\begin{equation}
		A : V \otimes V \otimes V \otimes V \to \mathbb{R}, \label{A}
	\end{equation}
	where $V$ is a finite-dimensional Euclidean space.
	
	In the tensor network approach, one first rewrites the partition function of
	a lattice model as a tensor network. This is usually simple to do explicitly,
	starting from the original lattice Hamiltonian. E.g.~in Appendix \ref{sec:ising} we rewrite the 2D Ising model on the square lattice in the form of Fig.~\ref{fig-intro1} using a four-tensor
	$A_{ijkl}$ where every index takes only two values, i.e. $\dim V = 2$.
	
	An RG transformation (or map) is a coarse-graining of the tensor network.
        It rewrites the tensor network in terms of new tensors, fewer in number than the original ones.
	The simplest rule defines a new tensor $T$ by contracting four $A$ tensors, as follows:
	\begin{equation}
		\myinclude[scale=0.5]{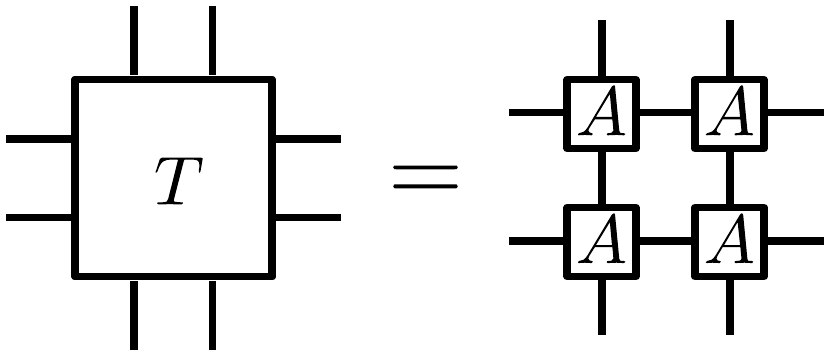}\,, 
		\label{A1}
	\end{equation}
	but more sophisticated rules can also be considered (see below). One then iterates this map and studies the resulting RG flow. 
	
	We see that $T$ is naturally defined on $W \otimes W \otimes W \otimes
	W$ where $W = V \otimes V$. It is in fact a general feature of tensor RG
	maps that they raise the index space dimension. In numerical calculations, reviewed in Appendix \ref{sec:prior}, it is customary to truncate $W$ to a subspace of the same
	dimension as $V$, chosen to minimize the truncation error. In contrast, in this paper we
	will not use any truncation. Our RG maps will preserve the partition
	function exactly. Our tensors will be defined in an infinite-dimensional
	real Hilbert space $V$ with a countable basis. For such $V$ 
	there exists a (non-unique) isomorphism\footnote{I.e.~a one-to-one isometric linear map. {See also footnote \ref{concreteJ}.}} between $V$ and $V \otimes V$. 
	Fixing some such isomorphism $J$, we define the final RG-transformed tensor $A'$ as
	\begin{equation}
		\myinclude[scale=0.5]{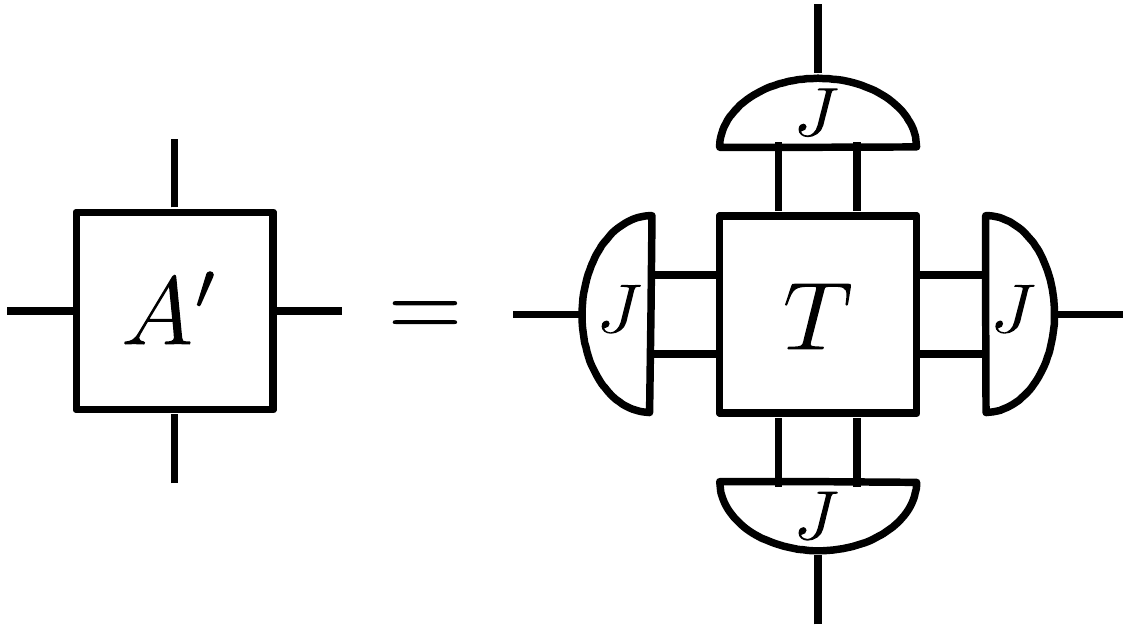}\label{XA}\,. 
	\end{equation}
	Now, tensors $A$ and $A'$ live on the same Hilbert space and can be compared. The RG fixed point equation in this setting reads
	\begin{equation}
		A'= \calN A ,\label{A1J} 
	\end{equation}
	where $\calN$ is a scalar. Whether this equation has a nontrivial solution may in general depend on the choice of $J$.
	
	It is instructive to compare tensor RG with the more traditional
	Wilson-Kadanoff RG in terms of Hamiltonians acting on spin variables.\footnote{See also \cite{Kadanoff} for such a comparison and a critical review of tensor RG by Leo Kadanoff and collaborators.}  There, starting
	with a lattice Hamiltonian $H$, a new Hamiltonian $H'$ is defined by the
	equation:
	\begin{equation}
		e^{H' [t]} = \sum_s B (s, t) e^{H [s]}\,,
		\label{WK}
	\end{equation}
	where $s$ are the original lattice variables, $t$ are the new (block-spin)
	variables, and $B (s, t)$ is a Kadanoff blocking function required to satisfy
	the condition $\sum_t B (s, t) = 1$ for any $s$, which guarantees that the
	partition function is preserved. The fixed point equation is $H' = H +
	Const.$
	
	In Wilson-Kadanoff RG we have a finite set of possible spin states $s_x$ on each lattice site $x$, replaced in tensor RG 
	by the infinite-dimensional Hilbert space $V$. This is a complication, but it is compensated by two
	advantages:
	
	\begin{enumerate} 
		\item Tensor RG keeps the interaction short-ranged: only
		nearest-neighbor tensors are linked by bonds in the network after any number of RG steps. 
		On the contrary, the new 
		Hamiltonian $H'$ will in general have infinite range after just one Wilson-Kadanoff RG step,
		even if the original Hamiltonian $H$ is short-ranged. 
		
		\item
		The definition
		of $A'$ is explicit, unlike that of $H'$. In fact, there is no
		known efficient algorithm to precisely evaluate the couplings of $H'$ in terms
		of those of $H$. Some approximate procedures do exist, but they cannot compute
		the fixed points and their associated critical exponents to very high and
		systematically improvable accuracy. For the same reason, a rigorous theory of nontrivial Wilson-Kadanoff fixed points has also never been constructed. 
	\end{enumerate}
	
	These advantages make tensor RG an attractive alternative to the
	Wilson-Kadanoff scheme. There is a lot of numerical evidence that tensor RG
	approaches perform very well, at least in 2D (see Appendix \ref{sec:prior}). In this
	paper we will initiate the rigorous theory of tensor RG.
	
	The rest of the paper is organized as follows. 
	As mentioned, our main goal is to discuss tensor RG near the high-temperature fixed point.
	This corresponds to the tensor $A_*$ with only one nonzero component $(A_*)_{0000}=1$. In section \ref{sec:stab} we will construct an appropriate RG map and prove that it converges to the high-temperature fixed point super-exponentially fast if we start in a small neighborhood of the fixed point.
	This is our main result  (Theorem \ref{main}). 
	The RG map will be somewhat more complicated than the basic RG map described above. As we will see, the basic RG map (referred to below as `type I') has eigenvalue 1 for a class of perturbations around $A_*$ in the linearized approximation. To deal with this problem we will introduce another RG map (called `type II'). The full RG step will consist of a type I and a type II RG maps applied one after the other. 
	
	In section \ref{sec:concl} we conclude and discuss perspectives for future rigorous work on the nontrivial fixed points. In Appendix \ref{sec:ising} we consider the 2D Ising model example. Appendix \ref{sec:prior} is a detailed review of prior work on tensor network RG.
	
	\section{Convergence to high-temperature fixed point}\label{sec:stab}
	
	In this section we will define our RG map and prove that if we start with
	a tensor sufficiently close to the high-temperature fixed point, then the sequence of tensors we get by iterating this map will converge to the high-temperature fixed point. Our RG map will be a composition of three steps which we will refer to as type 0, type I and type II. The type 0 step is just a gauge transformation, i.e.~a change of basis in the Hilbert space $V$. The type I step is the basic RG map described in the introduction. As already mentioned it's not contractive. Still, we keep it as a part of our RG map because it serves the useful purpose to add some specific structure to our tensor. The type II step, defined below, is the step that integrates out degrees of freedom in such a way that the resulting tensor has a smaller norm difference from $A_*$ than the tensor we started with.
	
	Let us fix a unit vector $e_0\in V$. The high-temperature fixed point for our RG map will be the tensor in which the
	only nonzero component is the $0000$ component and it has value $1$.
	We denote this tensor by 
	\begin{equation}
		\myinclude[scale=0.5]{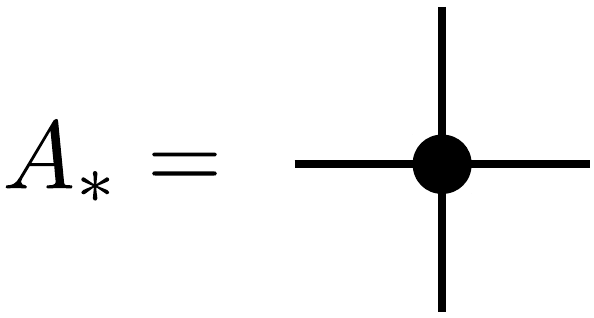}\,.
	\end{equation}
	Our tensor $A$ will always be a small, real, perturbation of this tensor.
	We will not have to assume that this perturbation preserves symmetries of the square lattice, i.e.~rotations by 90 degrees and reflections (see however Remark \ref{rem:refl}).
	
	{As mentioned we will work with infinite-dimensional tensors. Each index will run from 0 to $\infty$.} For tensors with any number of indices we will use the Hilbert-Schmidt norm (also known as the Frobenius norm). For example, if
	$A$ is a 4-index tensor then
	\begin{equation}
		\label{HSdef}
		\|A\| = \left( \sum_{i,j,k,l} A_{ijkl}^2 \right)^{1/2}\,.
	\end{equation}
	{The number of nonzero components may be finite or infinite, as long as the norm is finite. Finiteness of Hilbert-Schmidt norm is equivalent to the tensor defining a continuous linear mapping from $V\otimes V \otimes V \otimes V$ to $\mathbb{R}$ with $V=\ell_2(\mathbb Z_{\ge 0})$. One can think of a tensor abstractly as such a continuous linear mapping, or concretely as an infinite table of real numbers $A_{ijkl}$ such that the series in the r.h.s.~of \eqref{HSdef} converges.}

	We will make repeated use of the following property of the Hilbert-Schmidt norm. Let $A$
	and $B$ be tensors of any order. Let $C$ be a tensor formed by contracting
	some of the indices of $A$ with indices of $B$.
	Then the Cauchy-Schwarz inequality implies $\|C\| \le \|A\| \, \|B\|$. This can be iterated, e.g.~tensor $T$ defined by \eqref{A1a} has norm $\|T\|\le \|A\|^4$. Also it's easy to show that the result of such multiple contractions does not depend on the order in which they are performed, if the contracted tensors have finite norm. {All these elementary operations on infinite-dimensional Hilbert-Schmidt tensors work exactly as in finite dimension.}\footnote{For some other operations, such as taking traces or singular value decomposition, finiteness of Hilbert-Schmidt norm may not be enough in infinite dimension. We won't use those operations in this paper.}
	
	{Below we will also act on the legs of Hilbert-Schmidt tensors by bounded invertible linear operators. This preserves finiteness of the Hilbert-Schmidt norm. Acting by an isomorphism $J$ between $V$ and $V\otimes V$ is one example, and other examples (gauge transformations and disentanglers) will be introduced shortly.}
	
	We will often represent tensors with a diagram. When a bond is labelled with
	a $0$, it means that the index on that bond is set to $0$. When it is labelled
	with an $x$, it means that the index can range over all non-zero indices
	(if two bonds are labelled with $x$, the indices don't have to be identical).
	For unlabelled bonds the index ranges over all values.
	We think of this diagram as representing the tensor we obtain by setting all
	components that are not consistent with the bond labelling in the diagram
	to $0$. The norm of the diagram is then the Hilbert-Schmidt norm of this
	tensor. For example, the norm of diagram \eqref{fig_A3_comp} is\footnote{\label{conv1}We make the convention that $A_{ijkl}$ corresponds to the diagram with $i,j,k,l$ on the right, top, left and bottom legs, respectively.}
	\begin{equation}
		\left( \sum_{i \neq 0,j \neq 0} A_{ij00}^2 \right)^{1/2}\,.
	\end{equation}
	When we say that a diagram is $O(\epsilon^p)$ we mean that the norm of the
	tensor associated with the diagram is $O(\epsilon^p)$.

	We now introduce three different sets of assumptions A1, A2, A3 on the tensor $A$. Note that A3 is stronger than A2 which is stronger than A1.
	\begin{itemize}
		\item {\bf A1:}
		\begin{equation}
			\myinclude[scale=0.45]{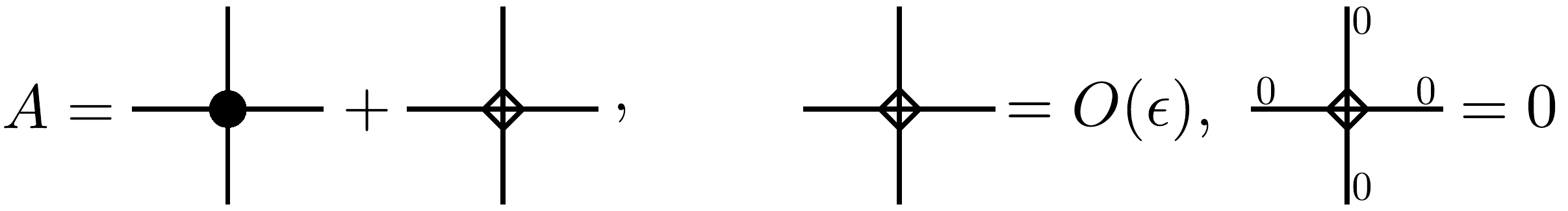}\,.
		\end{equation}
		\item {\bf A2:}
		\begin{equation}
			\myinclude[scale=0.45]{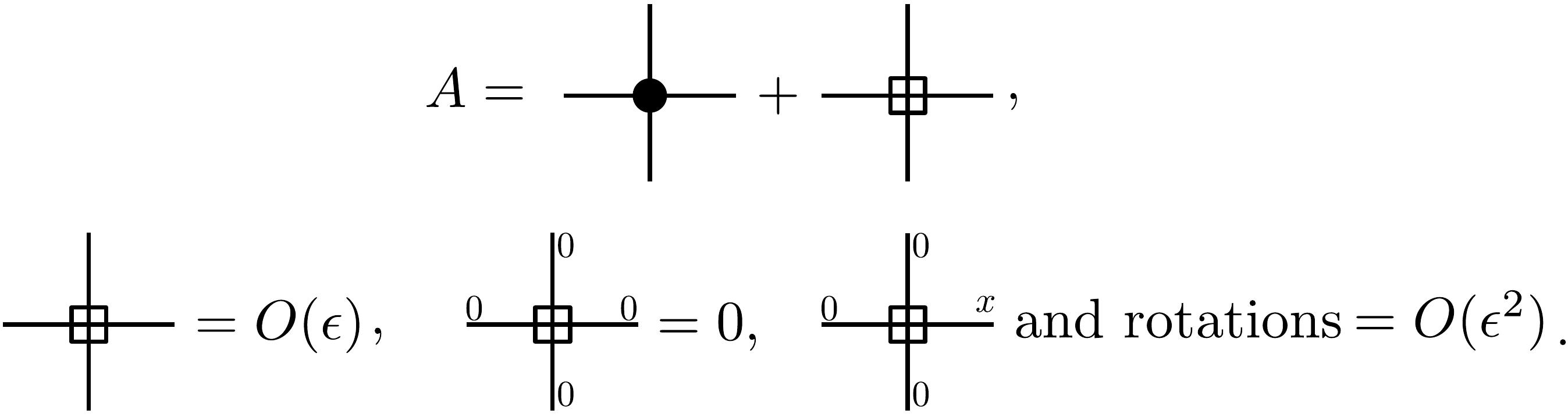}
		\end{equation}
		\item {\bf A3:}
		\begin{equation}
			\myinclude[scale=0.45]{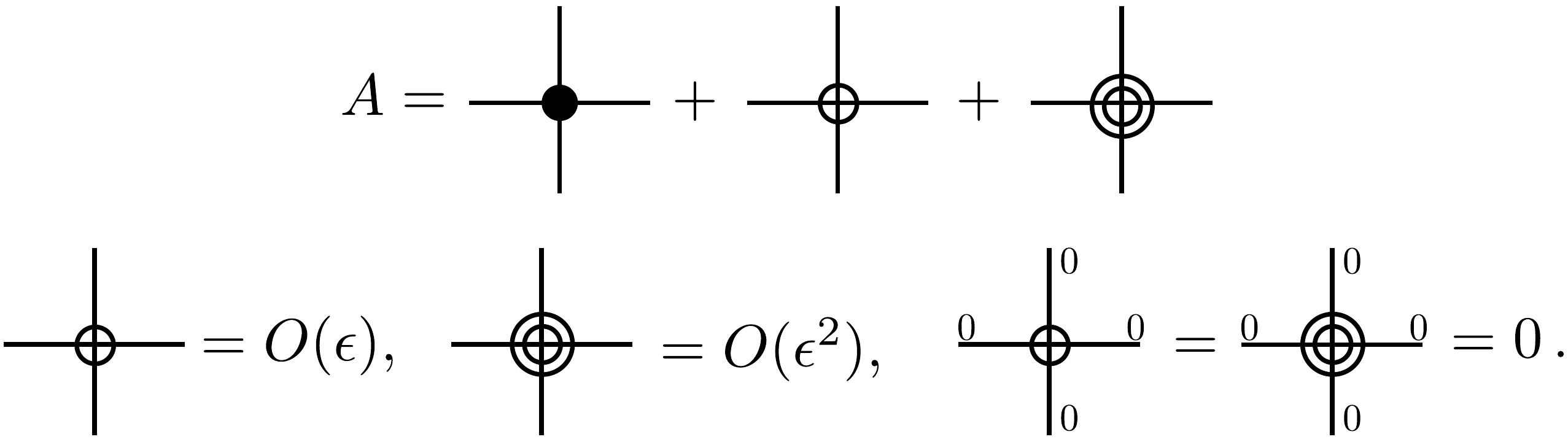} 
		\end{equation}
		In addition, A3 assumes that the only nonzero components of the one-circle tensor are the following "corner" diagram and its rotations:
		\begin{equation}
			\myinclude[scale=0.45]{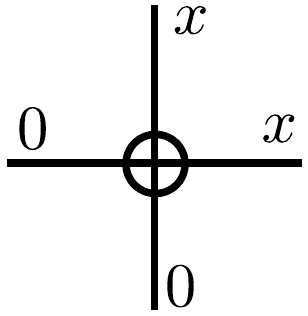}\,. 
			\label{fig_A3_comp}
		\end{equation}
This last assumption includes the assertion that components of the one-circle tensor with only one bond with a nonzero index are zero. 
	\end{itemize}

	We will prove three propositions. The first will show that if $A$ satisfies A1,
	then a gauge transformation will produce a tensor satisfying A2.
	The second proposition will show that if  $A$ satisfies A2,
	then the type I step will produce a tensor satisfying A3. Finally,
	the third proposition will show that if $A$ satisfies A3, then the
	type II step will produce a tensor that satisfies A1 with $\epsilon$ replaced by
	$\epsilon^{3/2}$.
	
	Each of A1, A2, A3 implies that tensor $A$ satisfies the normalization condition
	\beq
	A_{0000}=1\,. \label{normcond}
	\eeq
	This condition will in general be disturbed by our tensor manipulations.
	To restore it, we will \emph{normalize} the tensor at the end of each RG step, dividing it by a scalar:
	\beq
	A \mapsto A'=\calN^{-1} A, \quad \calN = A_{0000}.
	\eeq
	
	\subsection{Type 0}
	
	\begin{proposition} \label{prop:type0} If $A$ satisfies A1, then there is a gauge transformation
		such that the gauge-transformed $A$ satisfies A2,
		after normalization with $\calN=1+O(\eps^2)$.
	\end{proposition}
	
	\no {\bf Proof:}
	A gauge transformation is the insertion of $G G^{-1}$ into bonds in the
	network where $G$ is a bounded invertible operator. Since our network is
	not assumed invariant under rotations by 90 degrees, we will need to use different
	gauge transformations on the horizontal and vertical bonds, which we denote by
	$G_h$ and $G_v$. We will first use $G_h$ to reduce $A_{x000}$ and $A_{00x0}$ to $O(\eps^2)$ and we will then use $G_v$ to do the same for $A_{0x00}$ and $A_{000x}$ (recall footnote \ref{conv1}).
	
	Let $e_i\equiv |i\rangle$ ($i\ge 0$) be an orthonormal basis of $V$, with $e_0$ as the first element. 
	It is convenient to define a matrix $M$ as in the figure:
	\begin{equation}
		\myinclude[scale=0.45]{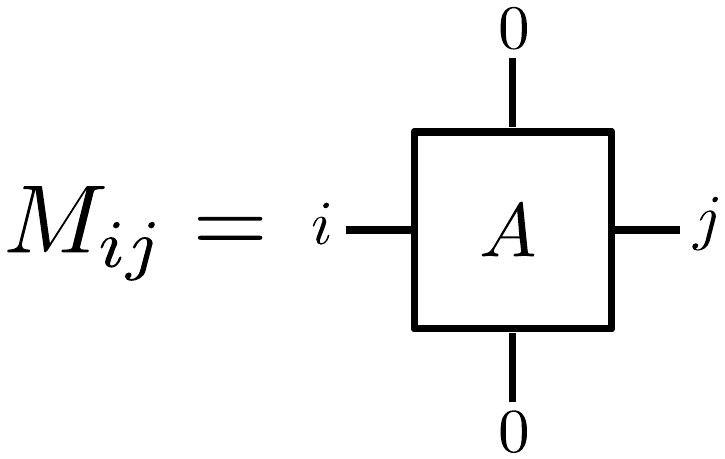}\,. 
	\end{equation}
	Note that $M_{00}=1$ by assumptions. We then let $l_x=M_{x0}$ and $r_x=M_{0x}$ ($x > 0$ by our conventions for the $x$ index).
	The hypothesis A1 implies that $\|l\|$ and $\|r\|$ are $O(\epsilon)$.
	(The norm is the $l^2$ norm.) Define two vectors by 
	\begin{equation}
		|l\rangle = \sum_x l_x |x\rangle, \quad \langle r| = \sum_x r_x \langle x|\,.
	\end{equation}

	We then define the linear operator $B:V\to V$ by
	\begin{equation}
		B = |l\rangle\langle 0|-|0\rangle\langle r|\,.
	\end{equation}
	Note that $B=O(\epsilon)$.
	Our gauge transformation for the horizontal bonds will be $G_h=\exp(B)$. 
	So $A$ is transformed to $\tilde{A} = G_h^{-1} A G_h$ :
	\begin{equation}
		\myinclude[scale=0.5]{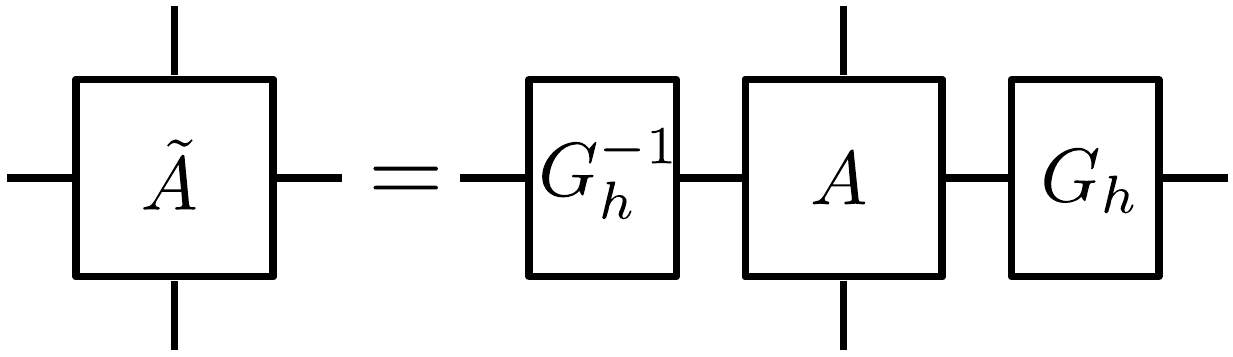}\ .
	\end{equation}
	Using the power series expansion of the exponential, we see that 
	\begin{equation}
		G_h = I +  B + O(\epsilon^2), \quad G_h^{-1} = I - B + O(\epsilon^2) \,.
	\end{equation}
	
	We claim that with this choice of $G_h$ we have $\tilde{A}_{x000}$ and $\tilde{A}_{00x0}$ both $O(\epsilon^2)$.
	Let $\tilde{M} = G_h^{-1} M G_h$. Then our claim is equivalent to $\tilde{M}_{0x}$ and $\tilde{M}_{x0}$ being both $O(\epsilon^2)$.
	We have
	\begin{eqnarray}
		\langle 0|G_h^{-1} M G_h |i\rangle   
		&=& \langle 0|(I - B + O(\epsilon^2)) M (I + B + O(\epsilon^2))|i\rangle  \nn\\[3pt]
		&=& \langle 0|M - B M + M B|i\rangle  + O(\epsilon^2)\nn \\[3pt]
		&=& M_{0i} + \langle r|M|i\rangle  - \langle 0|M|0\rangle  \langle r|i\rangle  + O(\epsilon^2)\,.
		\label{0i}
	\end{eqnarray}
Now let $i=x>0$. Since $\langle 0|M|0\rangle =1$ and $\langle r|x\rangle =r_x$, the term with these two factors
	cancels the $M_{0x}$. We are left with $\langle r|M|x\rangle $. The norm of this tensor (with a single index $x>0$)
	is $O(\epsilon^2)$ since the norm of $\langle r|$ is $O(\epsilon)$ and the $M$
	contributes another $O(\epsilon)$ since $M_{00}$ does not appear.
	Analogously $\langle x|G_h^{-1} M G_h |0\rangle =O(\eps^2)$. 
	
	Next we apply the gauge transformation on the vertical bonds to $\tilde{A}$. 
	The definition of $G_v$ is completely analogous to the horizontal case.
	But we need to check that this second gauge transformation does not undo what
	we accomplished with the first gauge transformation. So we need to
	show the following diagram is $O(\epsilon^2)$. 
	\begin{equation}
		\myinclude[scale=0.5]{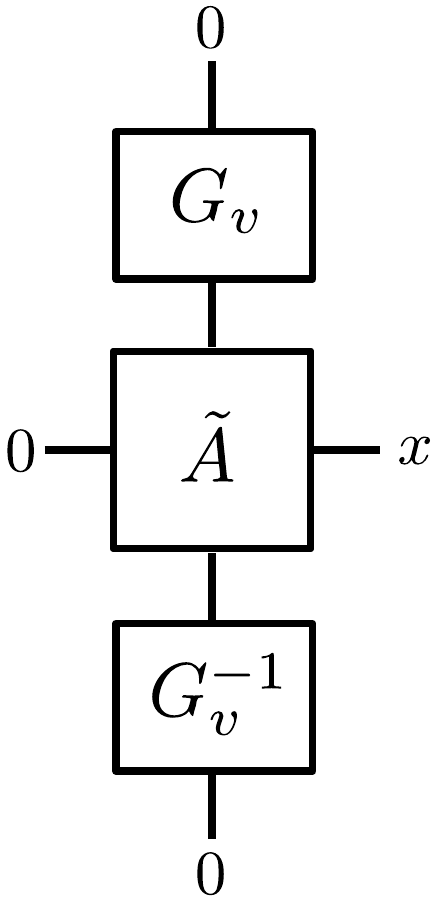}\ . 
	\end{equation}
	Note that in this diagram $\tilde{A}$ is $O(\epsilon)$ since it has at least
	one bond with a nonzero index. Since $G_v$ is of the form $I+O(\epsilon)$,
	the only term that could be $O(\epsilon)$ is when both $G_v$ and
	$G_v^{-1}$ are replaced by the identity. But this term is $O(\epsilon^2)$
	because of the first gauge transformation.
	
	The described transformations modify the 0000 component of the tensor making it $1+O(\eps^2)$. E.g.~we see that $\langle 0|G_h^{-1} M G_h |0\rangle =1+O(\eps^2)$ from \eqref{0i}. Dividing by $\calN=1+O(\eps^2)$ we get a tensor satisfying A2, including the normalization condition. \qed
	
	\begin{remark} \label{rem:refl}The given proof is valid whether or not $A$ preserves lattice symmetries. If $A$ is invariant under coordinate reflections, we have $r=l$, the operator $B$ is skew-symmetric and $G_h$ is an orthogonal transformation. 
	\end{remark}
	
	\subsection{Type I}\label{sec:typeI}
	
	The type I step is defined by Eqs. \eqref{A1} and \eqref{XA} in the introduction, which we copy here for convenience:
	\begin{equation}
		\myinclude[scale=0.5]{fig-intro2.pdf} \ , 
		\label{A1a}
	\end{equation}
	\begin{equation}
		\myinclude[scale=0.5]{fig-intro3a.pdf} .\label{XAa} 
	\end{equation}
	To fully specify it we must specify an isomorphism $J:V\stackrel{\sim}{\to}V\otimes V$. 
	We will assume that $J$ is such that:
	\beq
	J:e_0\mapsto e_0\otimes e_0\label{J1}\,.
	\eeq
	With this condition $A_*$ becomes a fixed point of the type I step. 
	Since an isomorphism preserves
	the inner product, \eqref{J1} implies that $J$ is an isomorphism of the orthogonal complements of $e_0$ and $e_0\otimes e_0$:
	\beq
	J:\ \text{span}(e_0)^\perp \stackrel{\sim}{\to} \text{span}(e_0\otimes e_0)^\perp\, .\label{J2}
	\eeq
	Clearly, this leaves $J$ vastly underdetermined.\footnote{\label{concreteJ}We stress however that $J$ can be specified very concretely if needed. For the simplest example, arrange the elements of the basis $e_i \otimes e_j$ of $V\otimes V$ in a single infinite list $\{f_k\}^\infty_{k=0}$ in the order of increasing $i+j$, and of increasing $i$ for equal $i+j$. Then the map $J$ mapping $e_k$ to $f_k$ does the job.} Proposition \ref{prop:typeI} below is valid for any such $J$.\footnote{We could also allow independent maps $J_h$ and $J_v$ on the horizontal and vertical bonds, but we will not need this.}
	
	\begin{proposition}[Type I RG step] \label{prop:typeI} Suppose $A$ satisfies A2. Define tensors $T$ and $A^\prime$
		by Eqs. \eqref{A1a} and \eqref{XAa}, where $J:V\stackrel{\sim}\to V\otimes V$ is any isomorphism satisfying \eqref{J1}.
		Then after normalizing with $\calN=1+O(\epsilon^4)$,
		$A^\prime$ satisfies A3.
	\end{proposition}
	
	\no {\bf Proof:}
	Assumption A2 decomposes $A$ into two tensors.  
	We insert this decomposition of $A$ into the definition of
	$T$ and expand $T$ to get 16 diagrams.
	Assumption A3 decomposes $A$ into a sum of three tensors, the high-temperature
	fixed point tensor and two more which we will refer to as
	the one-circle tensor and the two-circle tensor.
	
	Due to the condition \eqref{J1}, the zeroth order term reproduces the high-temperature fixed point
	tensor, i.e.
	\begin{equation}
		\myinclude[scale=0.45]{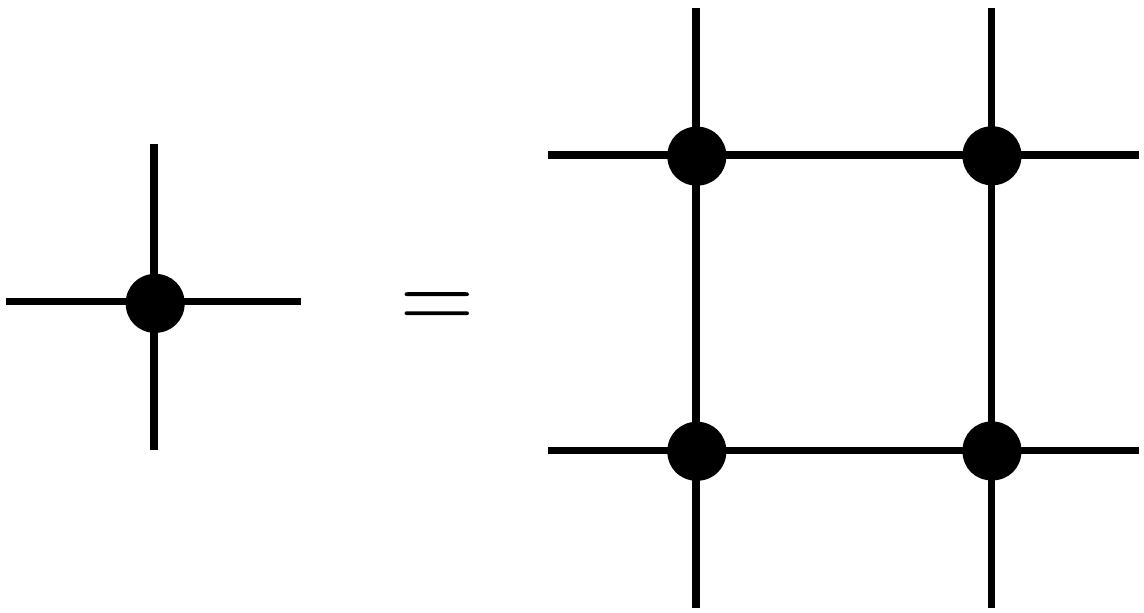}\,. 
	\end{equation}
	Here and below, we save space in the diagrams by leaving the contractions with the isomorphism $J$ implicit.
	
	We define the one-circle tensor by 
	\begin{equation}
		\myinclude[scale=0.45]{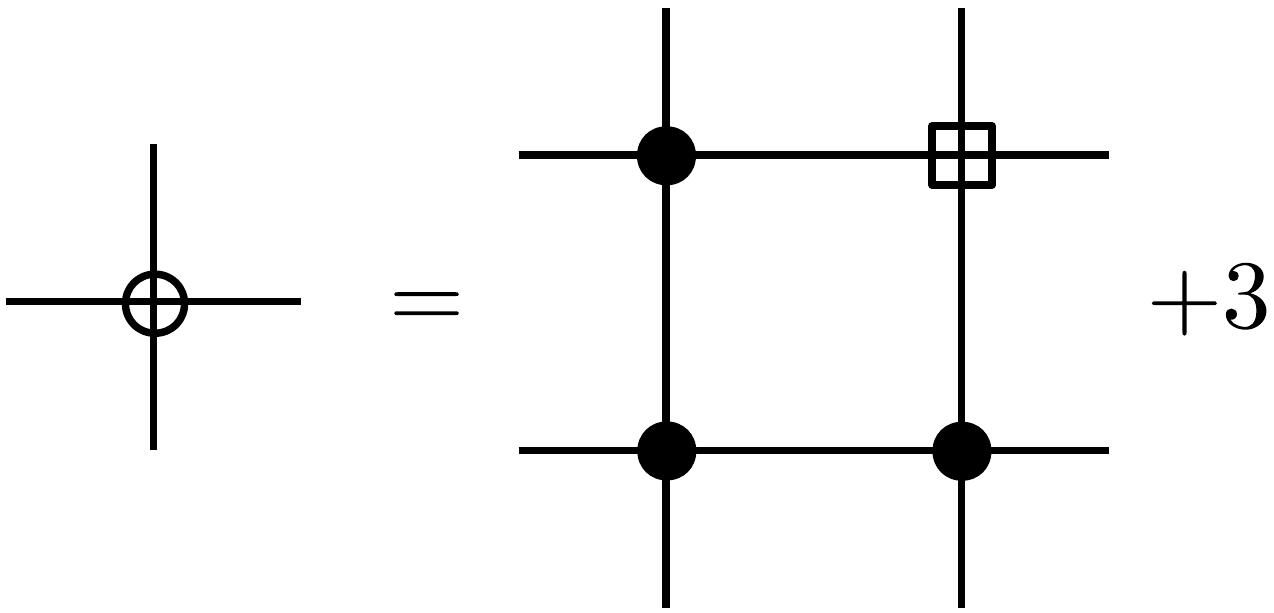}\,. 
	\end{equation}
The notation $+3$ is shorthand for the additional diagrams
        that we get by applying reflections about horizontal and vertical lines.
        Depending on the diagram, these reflections will generate $1$ or $3$
        additional diagrams which we will denote by $+1$ and $+3$, respectively.
        Recall however that the tensor $A$ itself is not required to have any symmetry.
        
	We then define the two-circle tensor to be the sum of the remaining 11
	diagrams. Clearly the one-circle tensor is $O(\epsilon)$, and 
	the two-circle tensor is $O(\epsilon^2)$. To verify the rest of
	A3, we note that the only nonzero components of the one-circle 
	tensor are those shown in the two subsequent figures (and rotations thereof):
	\begin{equation}
		\myinclude[scale=0.45]{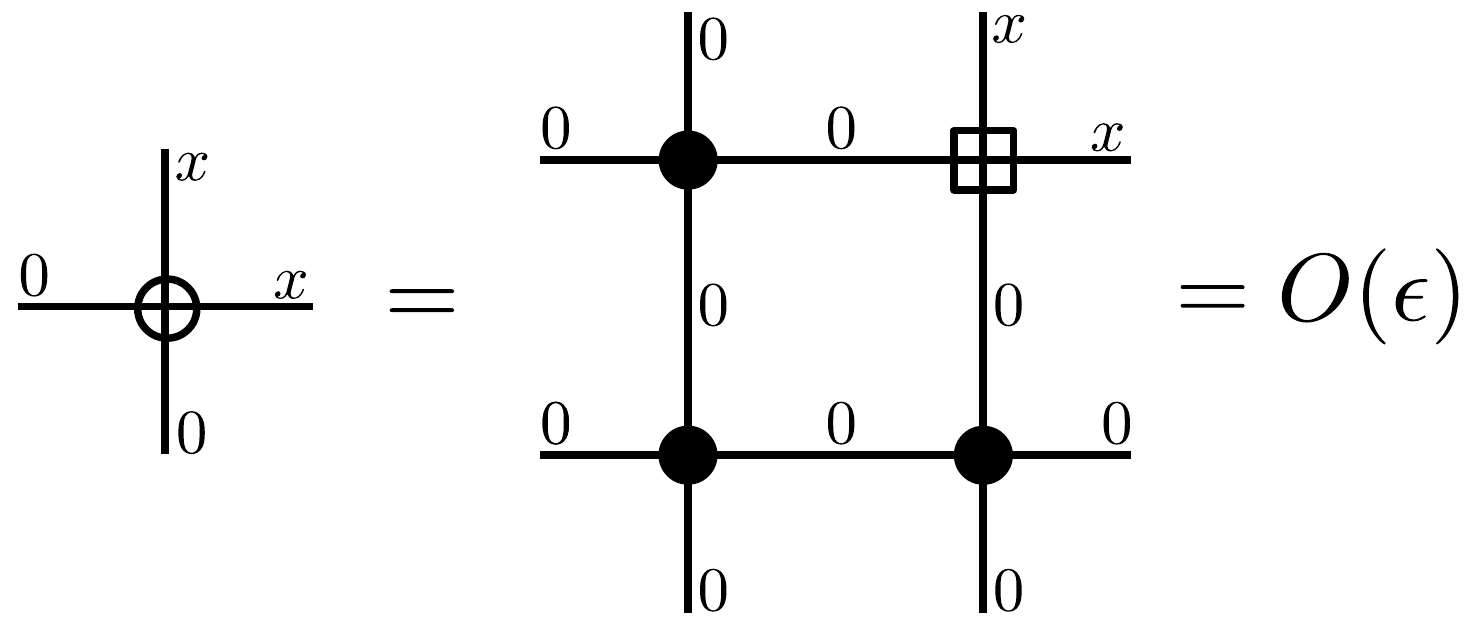}\ , 
		\label{fig_A3c}
	\end{equation}
	\begin{equation}
		\myinclude[scale=0.45]{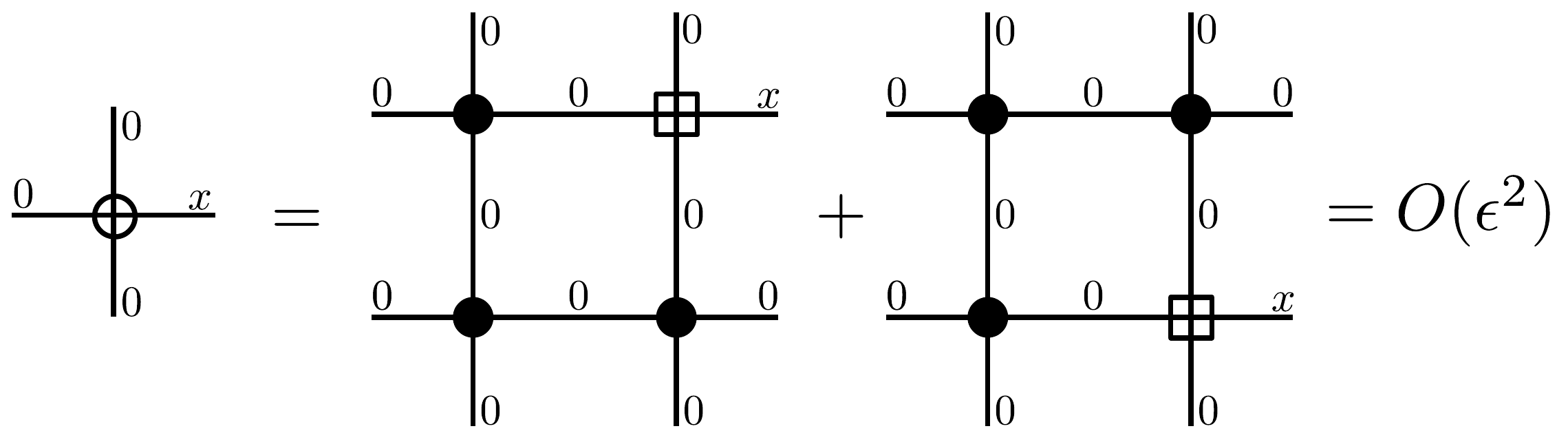}\ ,
		\label{fig_A3c1}
	\end{equation}
	where we used hypothesis A2 to estimate the order of diagrams in $\epsilon$. 
	Let us move terms in Eq.~\eqref{fig_A3c1} from the one-circle tensor
	to the two-circle tensor. The A3 assumptions for the one-circle tensor
	are now satisfied.
	
	At this point the 0000 component of the two circle-tensor is $O(\eps^4)$, coming from the diagrams
	\begin{equation}
		\myinclude[scale=0.4]{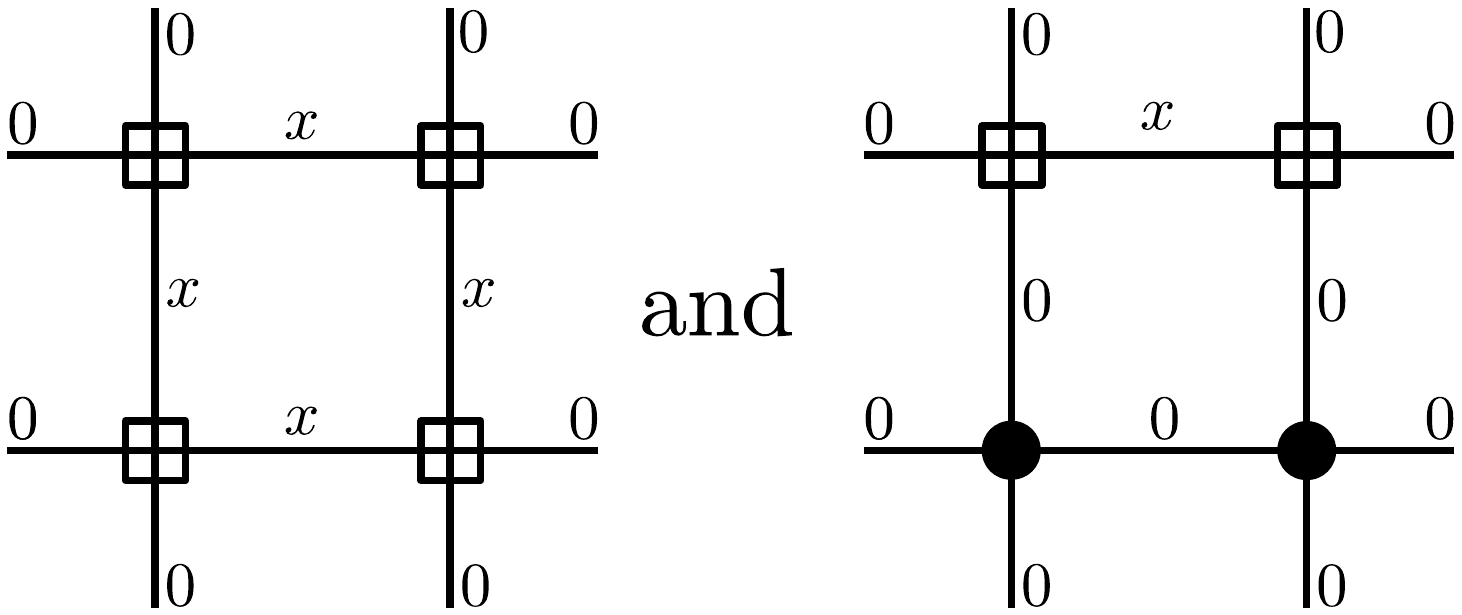} \,.
	\end{equation}
	Combining this component with the zero-order term, we can write the tensor $A'$ as
	\begin{equation}
		\myinclude[scale=0.5]{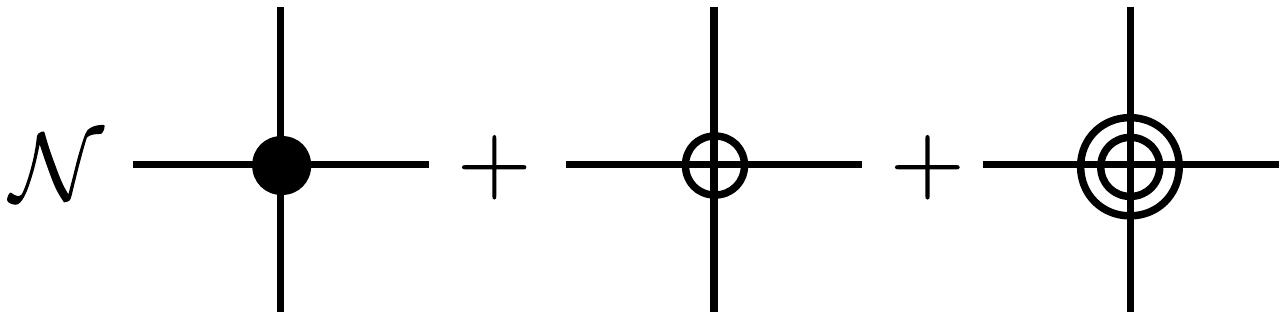} \,,
	\end{equation}
	where the one- and two-circle tensors satisfy the assumptions in A3 and $\calN=1+O(\eps^4)$ is a scalar. It remains to divide $A'$ by $\calN$ to satisfy A3.
	\qed
	
	\subsubsection{Eigenvalue 1 perturbations}
	
	This subsection may be skipped on the first reading as it is not strictly necessary to understand the rest of the construction. It explains a bit better why type I RG map by itself could not achieve our goal, motivating the need to introduce a more complicated type II map.
	
	We see from Proposition \ref{prop:typeI} that the type I RG step maps a tensor $A=A_*+\delta A$ with $\delta A=O(\eps)$ to $A'=A_*+\delta A'$ with $\delta A'=O(\eps)$. The $\delta A'$ has some extra structure which will be used in section \ref{sec:typeII}. Here we note that since both $\delta A$ and $\delta A'$ are $O(\eps)$, type I RG map iterates do not converge to $A_*$ superexponentially fast. One may inquire if there is at least exponential convergence, which would mean$\|\delta A'\| \le \lambda \|\delta A\|$	with some $\lambda<1$. As we will now show, even this weaker property does not hold. Namely there exist perturbations $\delta A$ which are eigenvectors with eigenvalue 1 in the linearized approximation, namely $\delta A'=\delta A+O(\eps^2)$.
	
	For the simplest example of such a perturbation, consider the tensor $\delta A$ which has only these four nonzero components, all equal to $\eps$:
	\begin{equation}
		\myinclude[scale=0.5]{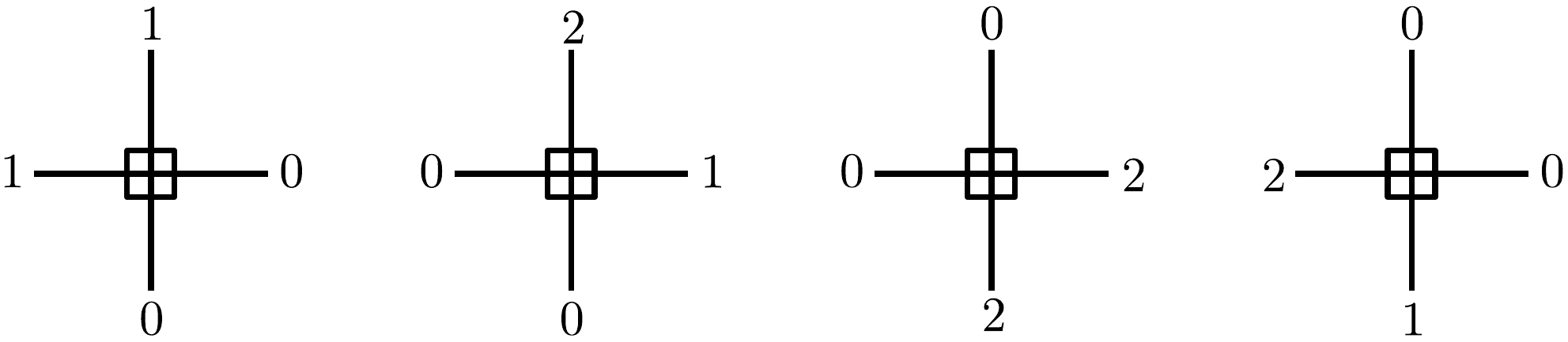}\,.
                \label{eig1}
	\end{equation}
	To understand better the structure of this tensor, note that it is invariant under parity transformations (flips about horizontal and vertical lines). Consider e.g.~flips about horizontal lines. Parity invariance means that the contraction of $A$ with 4 vectors $u, r, l, d\in V$ remains invariant if we interchange $u$ with $d$ and simultaneously change $r\to Pr$, $l\to Pl$ where $P:V\to V$ is a linear operator:  
	\beq
	\myinclude[scale=0.5]{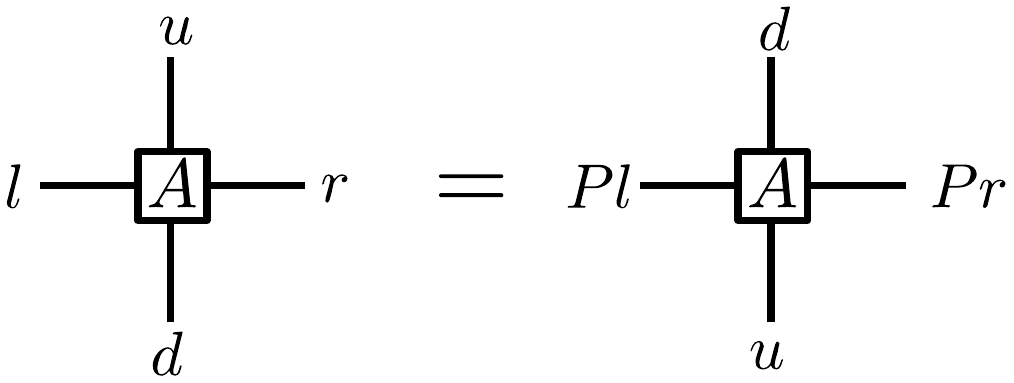}.
	\eeq
	Perturbation \eqref{eig1} of $A_*$ satisfies this property, as well as the analogous property for flips around vertical lines, for $P$ that leaves $e_0$ invariant and exchanges $e_1$ and $e_2$.
	
	All the components \eqref{eig1} are of corner type. Therefore, to linear order in $\eps$, the correction to tensor $T$ is given by the diagram in the top row of Eq.~\eqref{fig_A3c} and its rotations, namely:
	\beq
	\myinclude[scale=0.4]{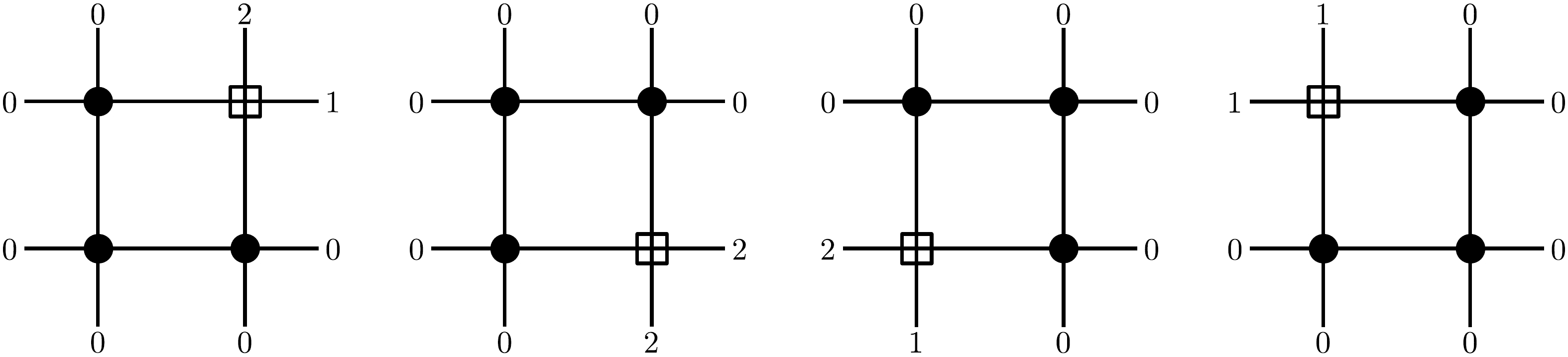}\,.
	\eeq
	The corresponding tensor elements are\footnote{The order of tensor components is according to the convention from footnote \ref{conv1}. We also make the convention for numbering basis elements of the tensor product $V\otimes V$: for vertical bonds the first factor in the tensor product corresponds to the left bond, while for horizontal bonds the first factor corresponds to the top bond.}  
	\beq
	T_{10,02,00,00}, \quad T_{02,00,00,02}, \quad T_{00,00,02,10},\quad T_{00,10,10,00}.
	\eeq
	To compute the corresponding $\delta A'$ we need to complete the definition
	of the $J$ isomorphism, compatibly with \eqref{J2}. We define:
	\beq
	J : e_1 \mapsto e_1 \otimes e_0 \equiv |10\rangle\,,  \qquad J : e_2 \mapsto e_0  \otimes e_2 \equiv |02\rangle \,.
	\eeq
	With this choice of $J$ we see that
	\begin{equation}
		\delta A'=  \delta A + O (\varepsilon^2)\,.
	\end{equation}
	A different choice of $J$ would amount to performing an orthogonal gauge transformation, leaving $\|\delta A'\|$ invariant.
	
	So perturbation $\delta A$ is left invariant (in the linearized approximation). At the same time, this perturbation can be considered "trivial" in the following sense. If one tries to construct a nonzero contraction of tensor components \eqref{eig1} and of the high-temperature tensor, one finds that the lines of indices 1,2 can never form a continuous line traversing the lattice. Rather, such lines "go around" single plaquettes, surrounded by 0 indices, as shown here:
	\beq
	\myinclude[scale=0.4]{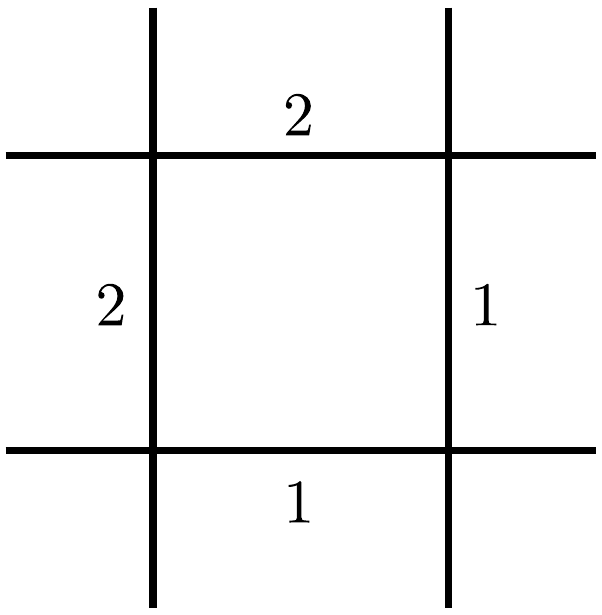}\,.
	\eeq
	Intuitively, this suggests that perturbations of this kind do not introduce long-range correlations in the tensor network. Although type I map fails to filter them, one hopes that a better RG map should be able to do so. Indeed, this will be achieved by type II map introduced below, which will have the property that $\delta A'=O(\eps^{3/2})$. 
	
	\begin{remark}\label{remCDL}
		Perturbation \eqref{eig1} is closely related to the so called corner-double-line (CDL) tensors from Appendix \ref{CDL}. 
		Filtering them away requires some ingenuity, as reviewed in Appendix \ref{sec:prior}. That is why the type II map will be somewhat complicated.
	\end{remark}
	
	\subsection{Type II}\label{sec:typeII}
	
	The type II RG step\footnote{This construction was inspired by the TNR algorithm by Evenbly and Vidal
		{\cite{Evenbly-Vidal}} from App.~\ref{TNRG}.} will be a composition of several smaller steps:
	\begin{enumerate}
		\item[II.1] Form the tensor $T$ by contracting four $A$ tensors (same as in type I RG map).
		
		\item[II.2] Introduce "disentanglers." The disentagler is a bounded invertible operator $R$ which acts on two indices. We
		insert $R R^{-1}=I$ into the two horizontal lines between $T$'s. We then combine $T$ with the $R^{-1}$ on its
		left and the $R$ on its right, and call the resulting tensor $S$:
		\begin{equation}
			\myinclude[scale=0.5]{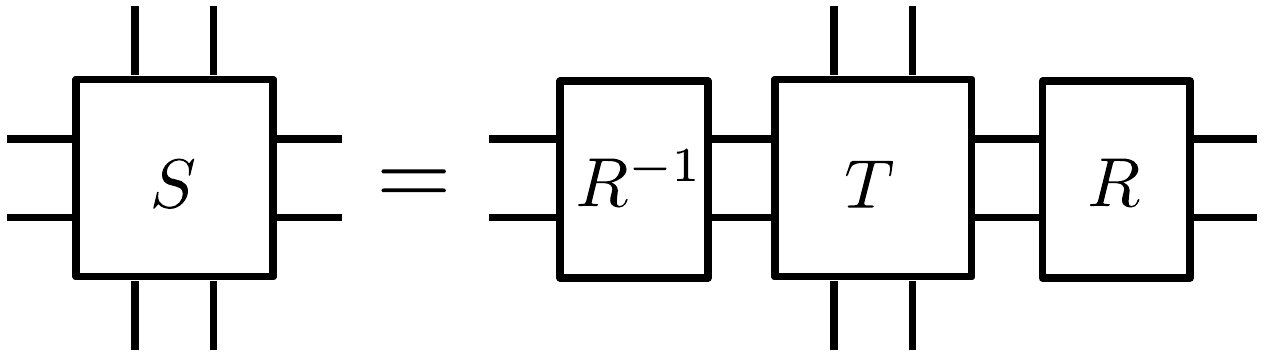}\,. 
			\label{def_S}
		\end{equation}
		This is similar to a gauge transformation except that the procedure operates on two indices.
		The choice of a good disentangler is a part of defining the RG map, and it will be discussed below.
		
		\item[II.3] Represent the tensor $S$ as a contraction of two tensors $S_u$ and $S_d$:
		\beq
		\myinclude[scale=0.5]{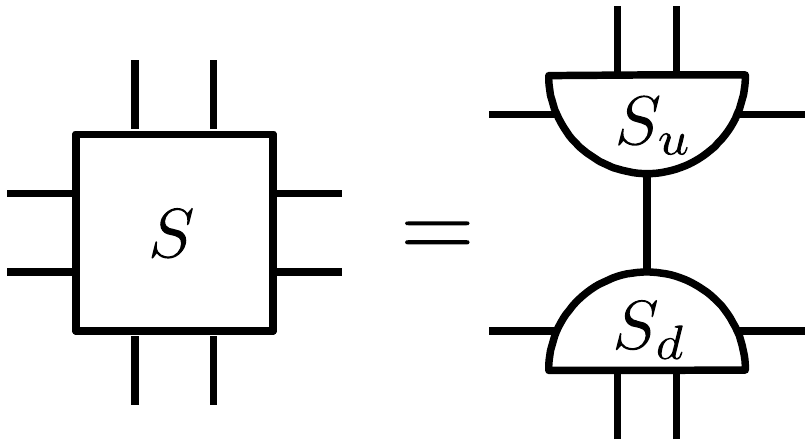} \,.
		\label{II3}
		\eeq
		\item[II.4] Define the tensor $U$ by contracting tensors $S_u$ and $S_d$ in the opposite order:
		\beq
		\myinclude[scale=0.5]{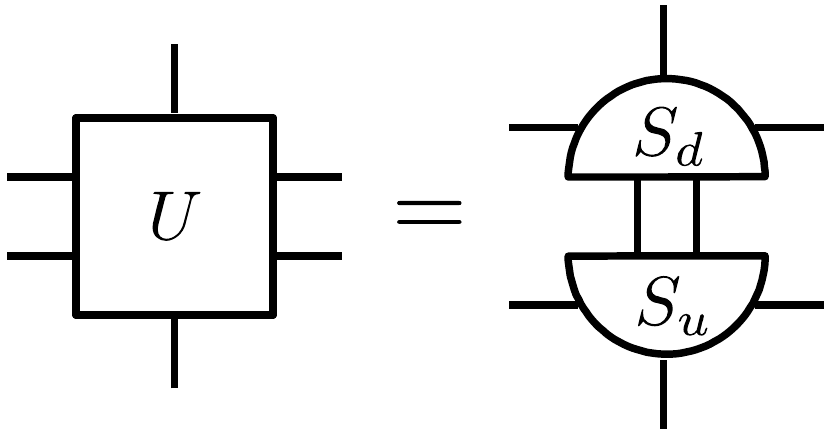} \,.
		\label{Udef}
		\eeq
		Since we are just changing the order of grouping the tensors, contracting $U$'s gives the same tensor network as before.
		\item[II.5] Define the final tensor $A'$ by contracting the tensor $U$ with two isomorphisms $J:V\to V\otimes V$:
		\beq
		\myinclude[scale=0.5]{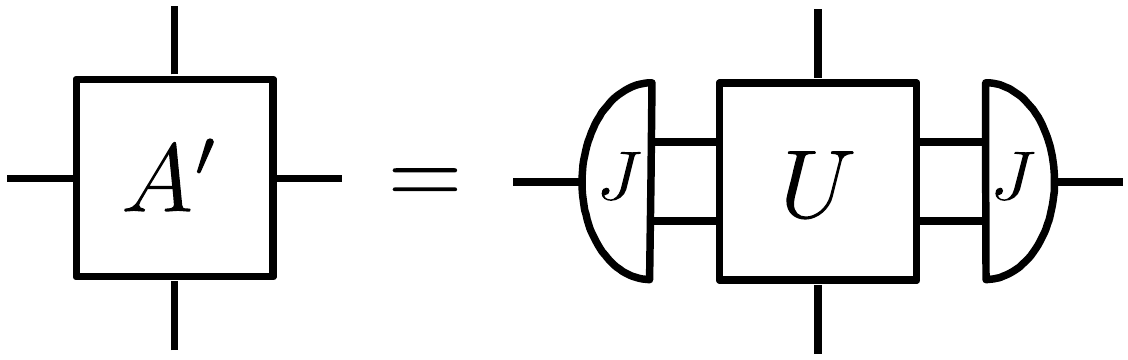} \,.
		\label{II5}
		\eeq
		For our purposes here any isomorphism satisfying \eqref{J1}, \eqref{J2} will do.
	\end{enumerate}
	
	\begin{proposition}[type II RG step] \label{prop:typeII}
		Suppose $A$ satisfies A3.
		There is a choice of the disentangler $R$ and of $S_u$ and $S_d$ in II.3
		so that the result $A^\prime$ of the 
		type II RG step satisfies A1 with $\epsilon$ replaced by $\epsilon^{3/2}$,
		after normalizing with $\calN=1+O(\eps^2)$. 
	\end{proposition}
	
	\no {\bf Proof:} Step II.1 consists in defining $T$ as before, Eq.~\eqref{A1}.
	Assumption A3 expresses $A$ as a sum of three tensors of orders
	$O(1),O(\epsilon),O(\epsilon^2)$. We insert this sum into each of the four
	copies of $A$ in the definition of $T$ and expand this to get $3^4$ diagrams.
	{We denote by $T_{\le 2}$ the sum of all diagrams which are $O(\epsilon^2)$:}
	\begin{equation}
		\myinclude[scale=0.4]{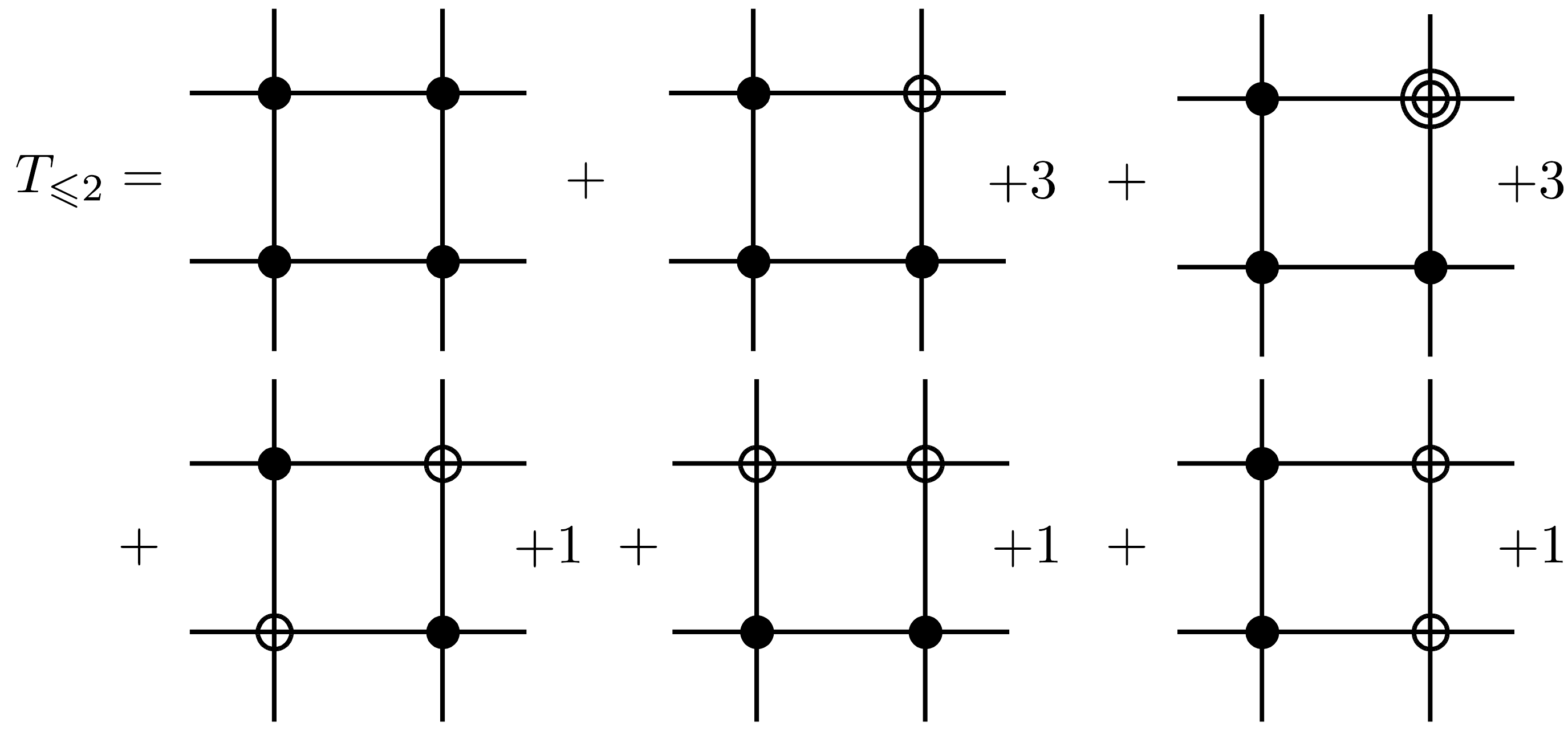}\ . 
	\end{equation}
     Recall that the notations $+3$ and $+1$ are shorthand for the additional diagrams
        that we get by applying reflections about horizontal and vertical lines.
        We denote by $T_>$ the sum of
	all the remaining diagrams, which are higher order than $O(\epsilon^2)$.
	So $T=T_{\le 2}+T_>$. 
	
	All diagrams in $T_{\le 2}$ except the last diagram have the property that the indices on
	the two internal vertical bonds are $0$. 
	We decompose the last diagram as follows:
	\begin{equation}
		\includegraphics[scale=0.4]{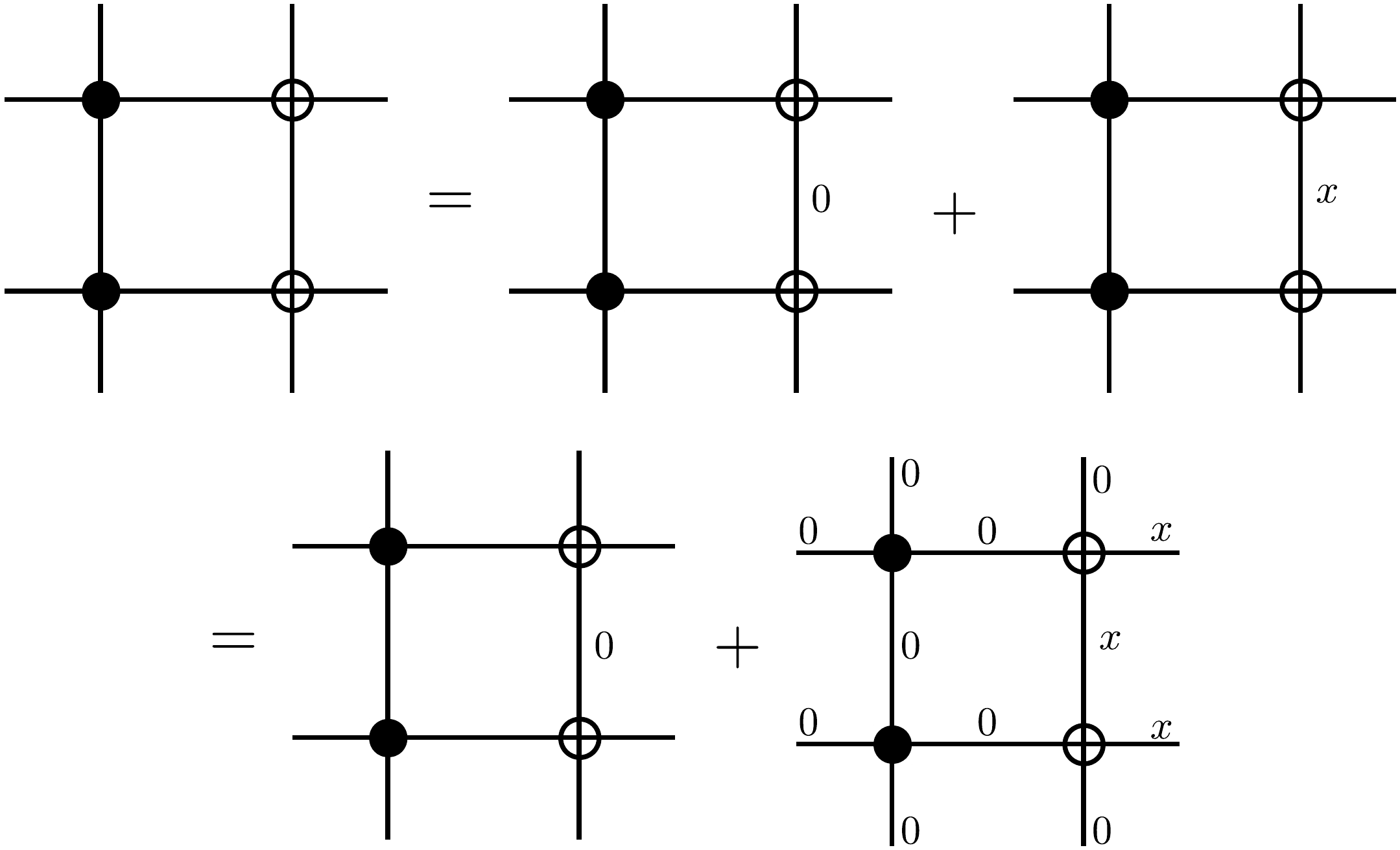} 
		\label{danger}
	\end{equation}
	We will refer to the last diagram (with all the bonds labelled) as the
	{\it dangerous diagram}. It is the only diagram in $T_{\le 2}$ 
	that does not have $0$'s on both of the vertical bonds. The dangerous diagram is $O(\epsilon^2)$. 
	In labelling its bonds we have made use of the
	assumption in A3 that the one-circle tensor only has corner nonzero components.
	
	We now proceed to Step II.2 which introduces disentanglers $R$ and defines the tensor $S$ as in \eqref{def_S}.
	We can write $S=R^{-1} T R$, with the understanding that it is the
	horizontal indices on $T$ being contracted in the products as indicated in the
	figure.
	The guiding principle to define the disentangler will be to remove the $O(\epsilon^2)$ part of the dangerous diagram.  We will see below why this is a good idea. The construction will be similar to how in the proof of Proposition \ref{prop:type0} we constructed gauge transformation $G_h$ removing the $O(\eps)$ part of $x$000 tensor components.
	
	For $i \neq 0$ and $j \neq 0$, define $l_{ij}$ and $r_{ij}$ as in the figure:
	\begin{equation}
		\myinclude[scale=0.4]{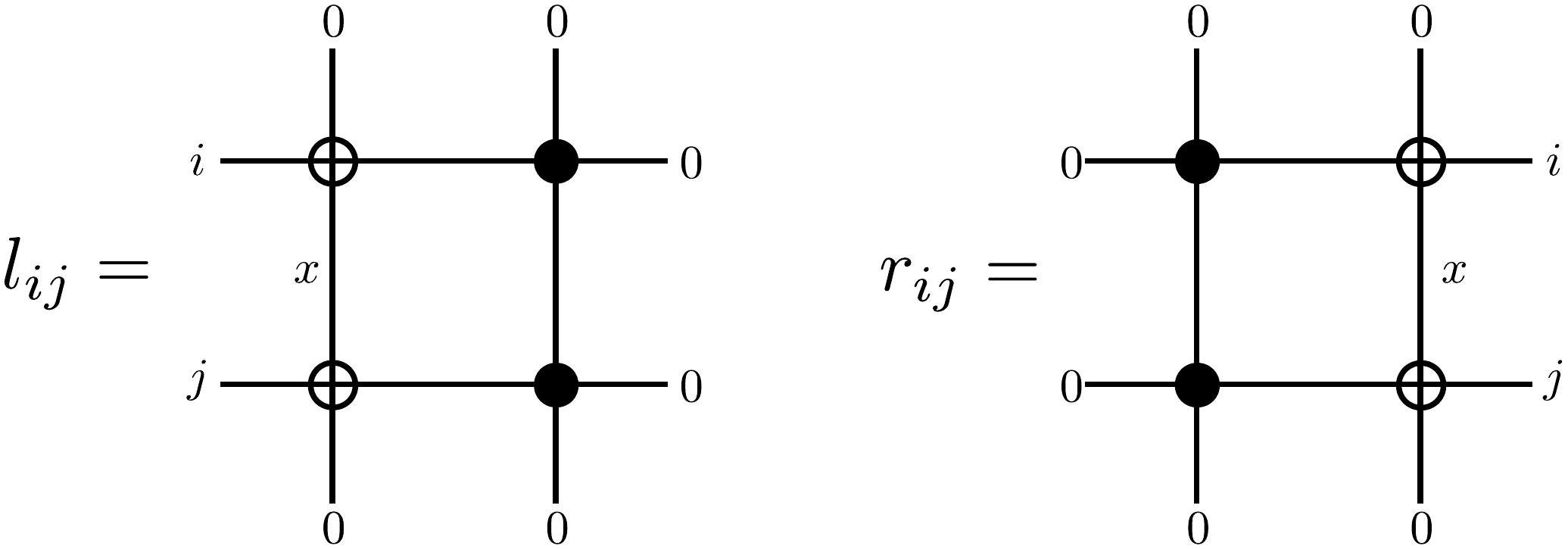}\ . 
	\end{equation}
	If one or both of $i$ and $j$ are zero, we define $l_{ij}=r_{ij}=0$. 
	Then we define 
	\begin{equation}
		|l\rangle  = \sum_{ij} l_{ij}|ij \rangle, \quad \langle r| = \sum_{ij} r_{ij} \langle i j|,
	\end{equation}
	where $|i j\rangle \equiv e_i\otimes e_j$.
	Note that these two vectors have norm $O(\epsilon^2)$,
	while in the case of the gauge transformation $G_h$ they had norm $O(\epsilon)$.
	We then define the linear operator $B$ from $V\otimes V$ to itself by
	\begin{equation}
		B =|l\rangle \langle 00|-|00\rangle\langle r|\,.
	\end{equation}
	Note that $B=O(\epsilon^2)$.
	We define $R=\exp(B)$.
 	So we have
	\begin{equation}
		\label{RRinv}
		R=I + B +C, \quad R^{-1} =I - B + \tilde C\,,
	\end{equation}
where $C,\tilde C =O(\epsilon^4)$.

	We now expand $S=R^{-1} T R$ using these expansions and $T=T_{\le 2}+T_>$.
	The key point here is that the two terms $-B T_0 + T_0 B$ 
	cancel the dangerous diagram and its reflection in \eqref{danger}.
	After these cancellations we see that we can write
	\beq
	S=S_{\le 2}+S_{>}\,,
	\eeq
	where $S_{\le 2}$ is the sum of the same diagrams as $T_{\le 2}$ \emph{except the dangerous diagram}:
	\begin{equation}
		\myinclude[scale=0.4]{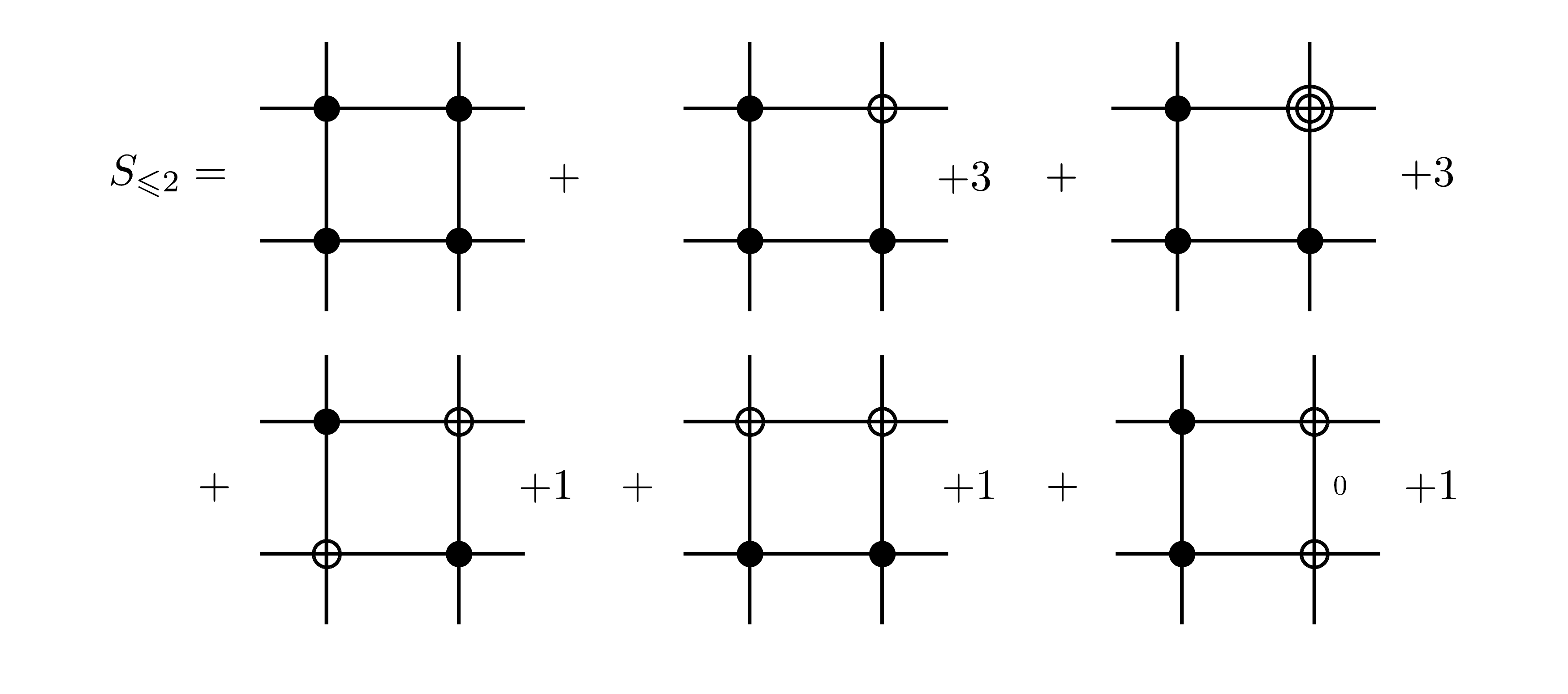}\ , \label{Sle2}
	\end{equation}
while $S_>$ includes
\begin{itemize}
	\item all diagrams in $R^{-1} T_> R$;
	\item all diagrams in {$R^{-1} T_{\le 2} R - T_{\le 2}$}
          except the two terms $-B T_0 + T_0 B$.
\end{itemize}

Note that all diagrams in $S_{\le 2}$ have zero indices on both internal vertical bonds. Note as well that $S_>=O(\eps^3)$. Achieving these two properties was the main point of the above construction. With $R$ and $R^{-1}$ expanded as in \eqref{RRinv}, $S_>$ consists of a finite number of diagrams.

	We now pass to step II.3. We will refer to the vertical bond in \eqref{II3} that connects $S_u$ and $S_d$ as the internal $S$-bond. We first define the half-disc tensor $S_u$ when the index on the internal $S$-bond is $0$.
	\begin{equation}
		\myinclude[scale=0.5]{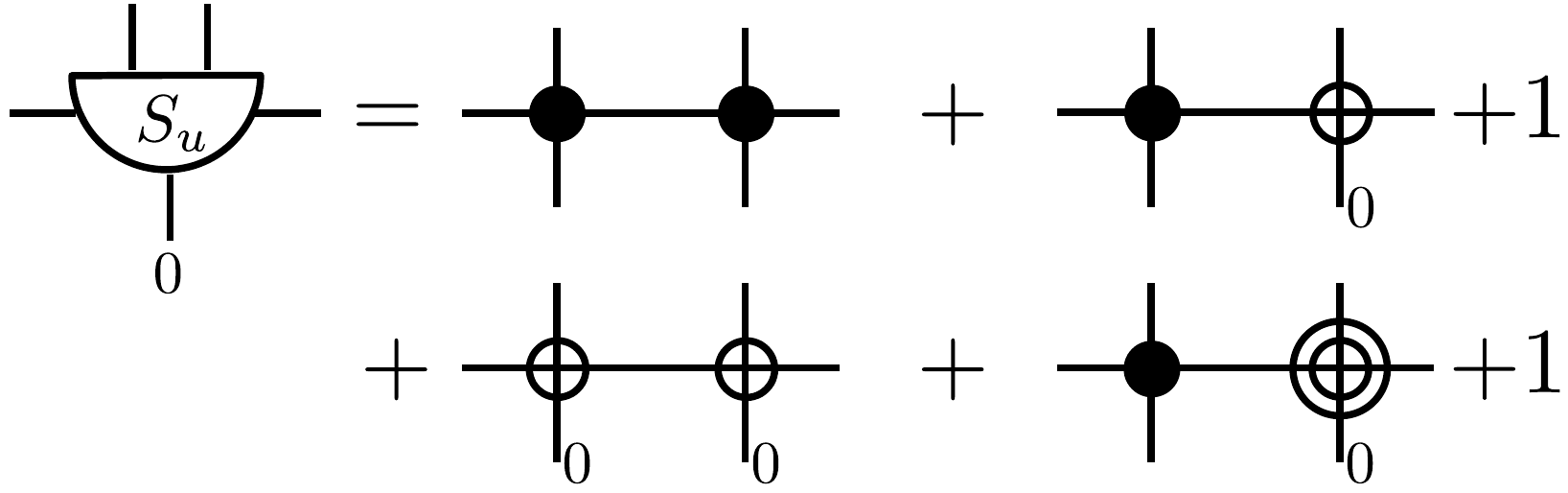}\ . 
		\label{Su}
	\end{equation}
	We define a similar half-disc $S_d$ by reflecting this figure about a horizontal
	line. When we contract the two half-discs with the internal $S$-bond restricted to $0$, we get:
	\begin{equation}
		\myinclude[scale=0.5]{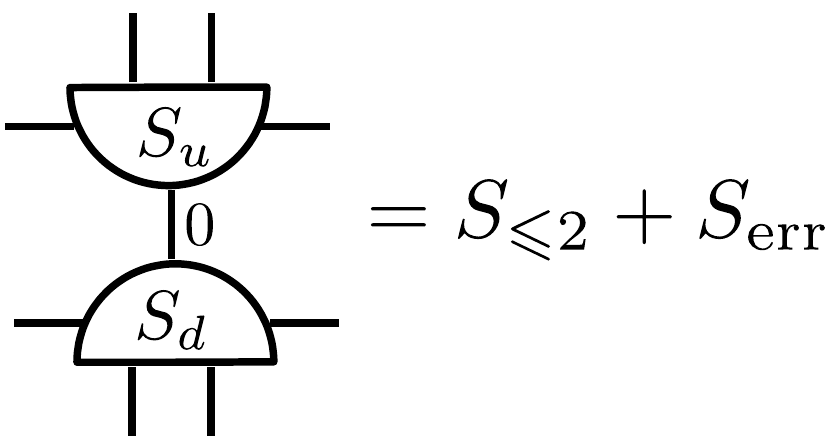} \ .
		\label{S_contract}
	\end{equation}
	Note in this contraction the index between the two half-disc is only
	summed over the single value $0$ (the other half-disc tensor components have not yet been defined). 
	The $S_{\rm err}$ term arises because the contraction in the figure above
	generates some diagrams that are not in $S_{\leq 2}$. It's easy to see that all
	such diagrams are $O(\epsilon^3)$. We now define
	\begin{equation}
		\Delta S = -S_{\rm err} + S_>=O(\eps^3)\,.
	\end{equation}
  
        {We will now show that we can define the rest of the tensor components of these half-disc tensors for indices on the internal $S$-bond other than $0$ so that their contraction when the internal $S$-bond ranges over all indices other than $0$ reproduces $\Delta S$, and so that the full half-discs are $O(\epsilon^{3/2})$.} In numerical tensor network RG literature, a typical method of representing a tensor as a contraction of two tensors involves singular value decomposition (SVD). Here we will not use SVD, rather we will use the fact that we know precisely which diagrams appear in $\Delta S$. Each such diagram will be naturally written as a contraction of two half-disc tensors (the upper and the lower half of the diagram). Since each diagram is $O(\epsilon^3)$, by moving factors of $\epsilon^{1/2}$ we will be able to make each of the half-discs $O(\epsilon^{3/2})$.

        The Hilbert space for the internal $S$-bond will have the form $\bigoplus_{n=0}^N W_n$. $W_0$ is one-dimensional and corresponds to the index $0$ for the bond. The rest of the $W_n$ are infinite dimensional (with a countable basis). So this direct sum is isomorphic to the original Hilbert space $V$. Let $N$ be the number of diagrams in $\Delta S$. For $n=1,2,\ldots, N$, the contraction of $S_u$ and $S_d$ with the internal $S$-bond restricted to $W_n$ will equal the $n$th diagram in $\Delta S$. 
          
        We explain how to define $S_u$ and $S_d$ by considering examples of specific diagrams. Recall that we expanded $R$ and $R^{-1}$ as in \eqref{RRinv}. We first consider a couple of diagrams in which $R$ and $R^{-1}$ have both been replaced by just $I$. 
        	
        	Our first example is:
        	\begin{equation}
        		\myinclude[scale=0.5]{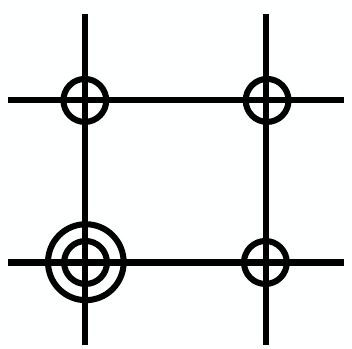} .
        	\end{equation}
        	This diagram was in $T_{> 2}$ and it enters also in $S_{> 2}$ and hence into $\Delta S$. It is represented as a contraction of $S_u$ and $S_d$ tensors given by the upper and lower halves of the diagram:
        	\begin{equation}
        		\myinclude[scale=0.5]{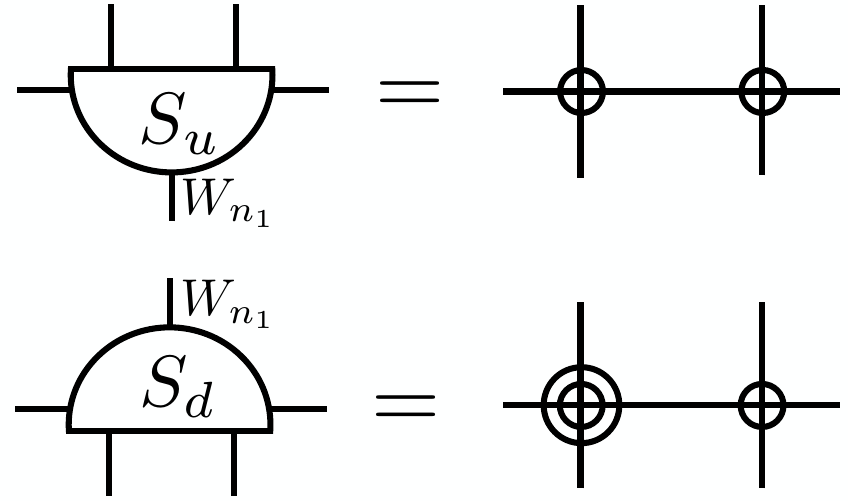}\ .
        	\end{equation}
        The vertical bond space $W_{n_1}$ (where $n_1$ is the number of this diagram) is isomorphic to $V\otimes V$. Note that in this example $S_u=O(\eps^2)$ and $S_d=O(\eps^3)$, i.e., even smaller than $O(\eps^{3/2})$ declared above.
        	
	For the second example consider the following diagram from $T_{> 2}$:
	\begin{equation}
		\myinclude[scale=0.4]{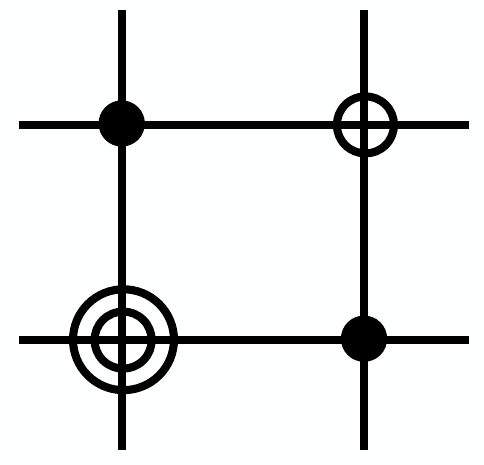} .
	\end{equation}
  Like in the previous example, we will represent this diagram as a contraction of $S_u$ and $S_d$ tensors given by the upper and lower halves of the diagram. The difference from the previous example is that we need to rescale the tensors by $\eps^{1/2}$ factors to make them both $O(\eps^{3/2})$:
	\begin{equation}
		\myinclude[scale=0.5]{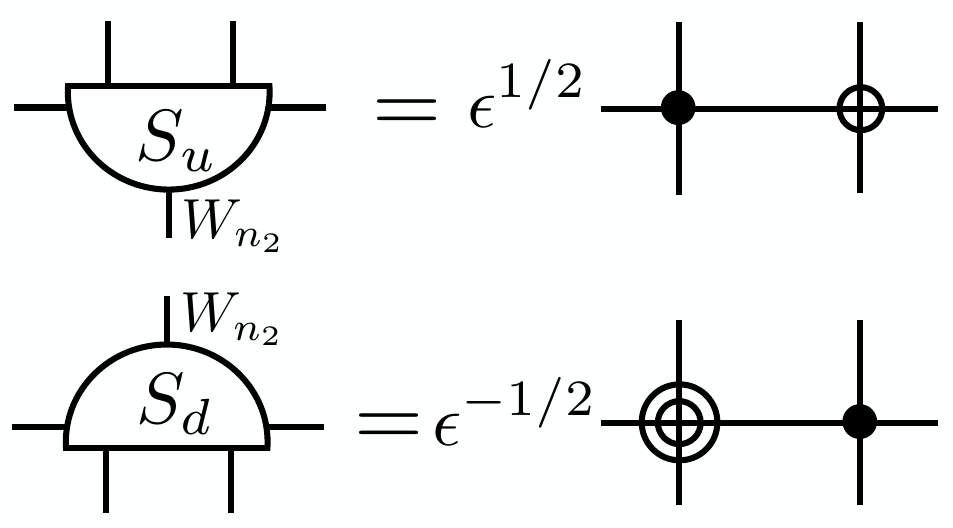}\ .
	\end{equation}
The vertical bond space $W_{n_2}$ is again isomorphic to $V\otimes V$.\footnote{\label{note:truncate}Because of how the high-temperature tensor $A_*$ enters into $S_u$ and $S_d$, the only nonzero term in the contraction is when both indices on the vertical legs are 0. We could therefore additionally restrict $W_{n_2}$ to the one-dimensional space spanned by $e_0\otimes e_0$. This is a minor detail which does not change the scaling of $S_u$ and $S_d$ with $\eps$.}

All the other diagrams in $S_{\rm err}$, and the diagrams in $S_{> 2}$ where $R$ and $R^{-1}$ have both been replaced by just $I$ are handled similarly to the above two examples.

For diagrams in which $R$ or $R^{-1}$ are not just replaced by $I$, we will need a small lemma representing $B$ and $C,\tilde C$ as a contraction of two tensors. This will allow us to cut the corresponding diagrams in two.
	
	\begin{lemma} \label{lem:Repr} 1. Tensor $B$ can be represented as a contraction of two three-leg tensors
		\begin{equation}
				\myinclude[scale=0.4]{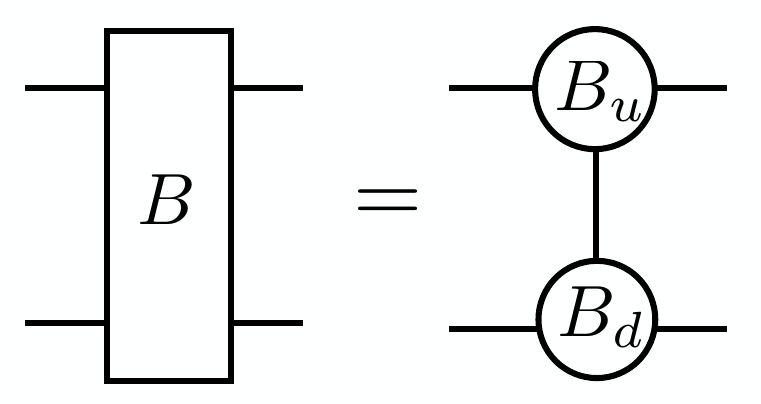} \ ,
			\label{eq:Brepr}
		\end{equation}
		where $B_{u,d}=O(\epsilon)$.\\
		2. Tensor $C$ in \eqref{RRinv} can also be represented as a contraction of two three-leg tensors
		\begin{equation}
			\myinclude[scale=0.4]{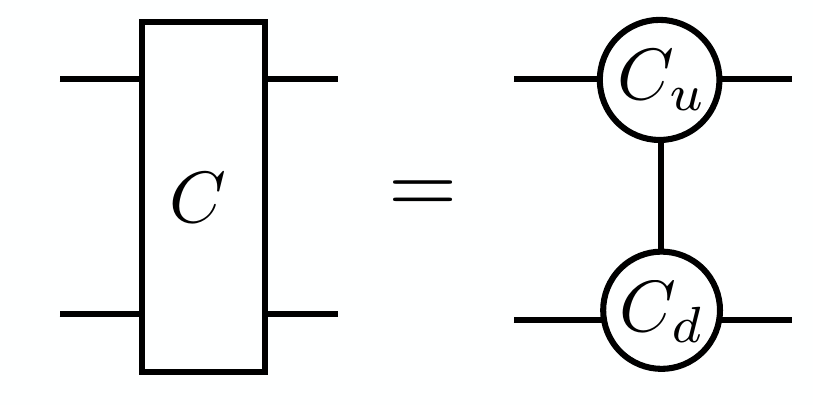} \ ,
			\label{eq:Rrepr}
			\end{equation}
			where $C_{u,d}=O(\eps^2)$. Tensor $\tilde C$ can also be represented in this form, replacing $C_{u,d}$ by $\tilde C_{u,d}=O(\eps^2)$.
		\end{lemma}
	
		\no {\bf Proof:}
We denote by $V_B$ and $V_C$ the Hilbert spaces for the vertical bonds joining $B$ and $C$ tensors in \eqref{eq:Brepr} and \eqref{eq:Rrepr}. We will take $V_B$ to be spanned by two copies of nonzero basis elements of $V$, we index these copies by $(x,0)$ and $(0,x)$ where $x\ne0$ is in $V$. For Part 1, we assign (note a minus sign in the last equation):\footnote{We thank Nikolay Ebel for pointing out a mistake in the previous version of the proof.} 
 \begin{equation}
 	\myinclude[scale=0.5]{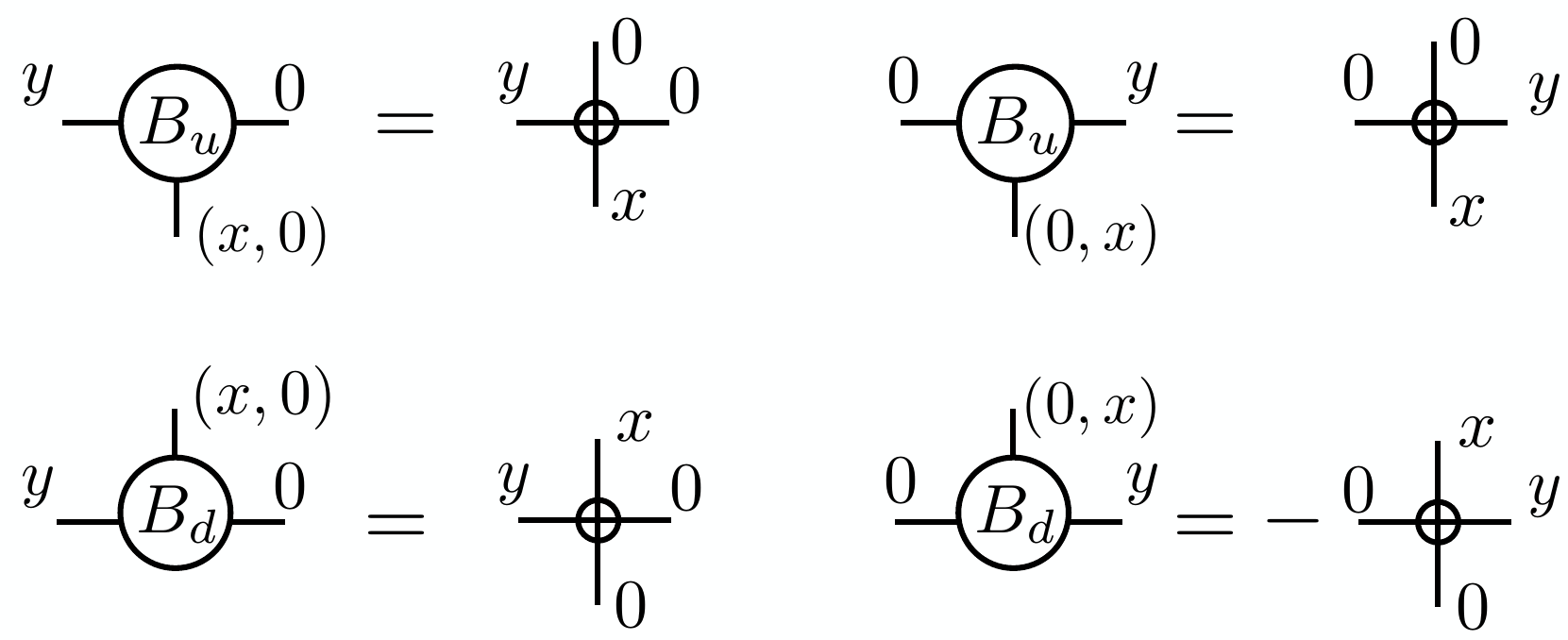} \ .
 \end{equation}

 For Part 2, we note that $C=\exp(B)-B-I$ can be represented in the form \eqref{eq:Rrepr} with
 \begin{equation}
 	\myinclude[scale=0.5]{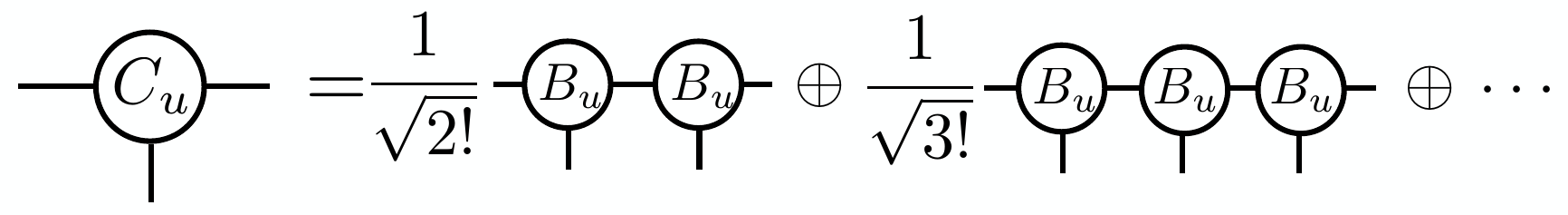} \ ,
 \end{equation}
and an analogous expression for $C_d$. It is a direct sum since every term has its own vertical Hilbert space $(V_B)^{\otimes n}$ for the term with $n$ $B_u$-tensors. The vertical Hilbert space of the $C_u$-tensor is the direct sum of Hilbert spaces of individual terms:
$V_C=\bigoplus_{n=2}^\infty (V_B)^{\otimes n}$. The norm of $C_{u,d}$ is bounded by $\sum_{n=2}^\infty \|B_u\|^n/\sqrt{n!}=O(\eps^2)$. 

To represent $\tilde C = \exp(-B)+B-I$ note that $-B$ can also be represented as in \eqref{eq:Brepr} by changing $B_u\to -B_u$. \qed
\vspace{0.5cm}

We can now consider the remaining diagrams in $S_{> 2}$. They are treated using the same main ideas as in the first two examples: cut the diagram in two and move factors of $\eps^{1/2}$ if needed. Since all the diagrams are $O(\eps^3)$, the resulting tensors $S_u,S_d=O(\eps^{3/2})$. We will consider just one example in detail.

Take the following diagram where $R^{-1}$ is replaced by $I$ and $R$ is replaced by $B$ (which we represent as a contraction of $B_u$ and $B_d$):
	\begin{equation}
		\myinclude[scale=0.4]{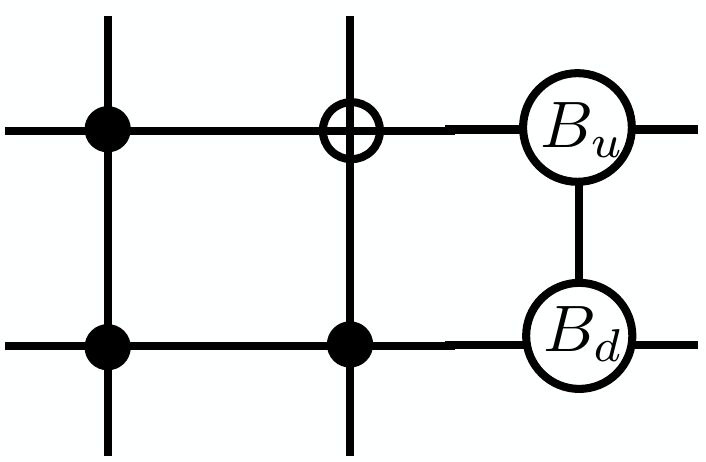}\,. 
	\end{equation}
       We represent it as a contraction of $S_u$ and $S_d$ tensors given by cutting the diagram in two and moving $\eps^{1/2}$ factors:
\begin{equation}
	\myinclude[scale=0.4]{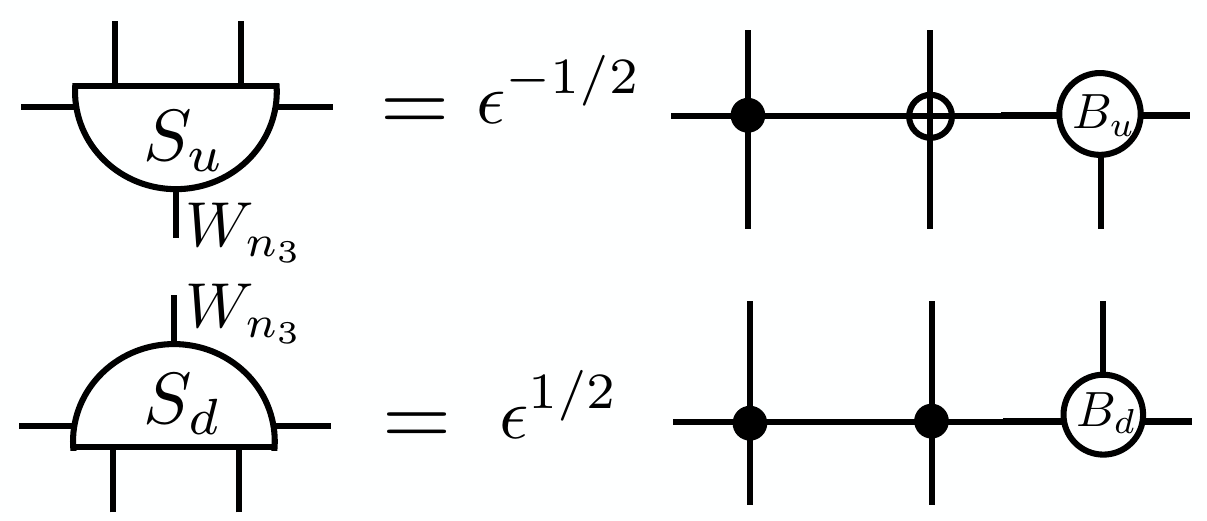}\ .
\end{equation}
In this case $W_{n_3}=V\otimes V \otimes V_B$.\footnote{And like in footnote \ref{note:truncate} it can be truncated setting the first two indices to 0, without changing the contraction.}
The other diagrams in which either $R$ or $R^{-1}$ is not replaced by $I$ are handled similarly. 

{By moving factors of $\epsilon^{1/2}$ and using the lemma where needed, we can represent each diagram in $\Delta S$ as a contraction of two half-disc tensors (the upper and the lower half of the diagram) where the index for the internal $S$-bond ranges over $W_n$ for the $n$th diagram, and each of the half-discs is $O(\epsilon^{3/2})$. Thus we have
	\begin{equation}
		\myinclude[scale=0.5]{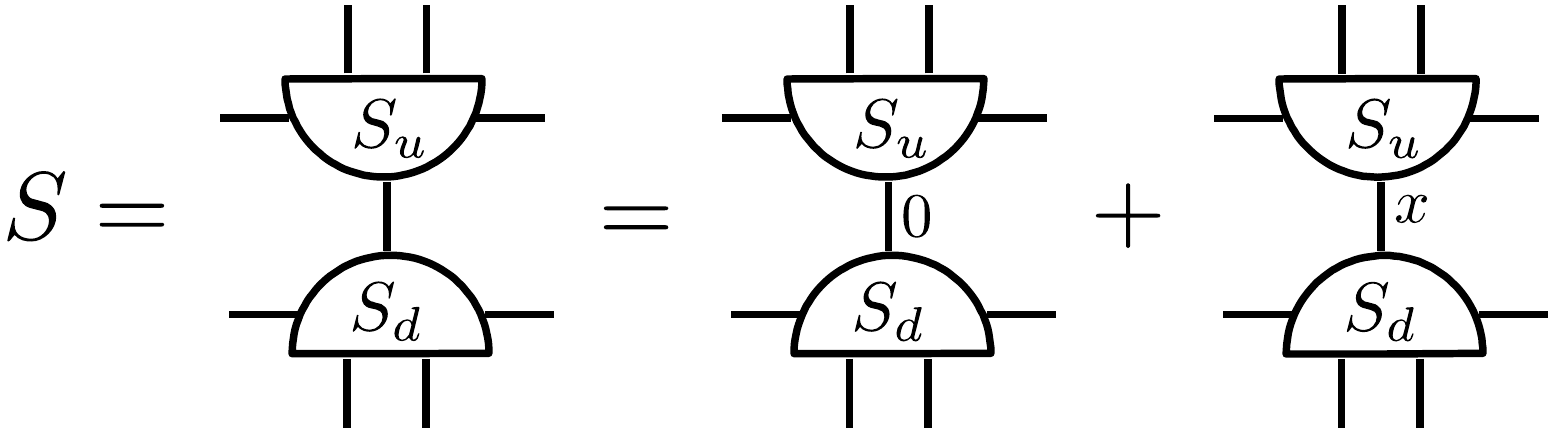}\,, 
	\end{equation}
	where the first contraction in the r.h.s.~is \eqref{S_contract} and the second contraction is the representation of $\Delta S$, where the index marked with $x$ numbers the orthonormal basis of the direct sum $\bigoplus_{n=1}^N W_n$. The half-disc tensors are $O(1)$ in the first contraction and $O(\eps^{3/2})$ in the second contraction. This ends step II.3.
        }
	
	Let us now move to step II.4 of our RG map. In it we change how we group terms, defining the $U$ tensor as in \eqref{Udef}. We have:
	\begin{equation}
		\myinclude[scale=0.5]{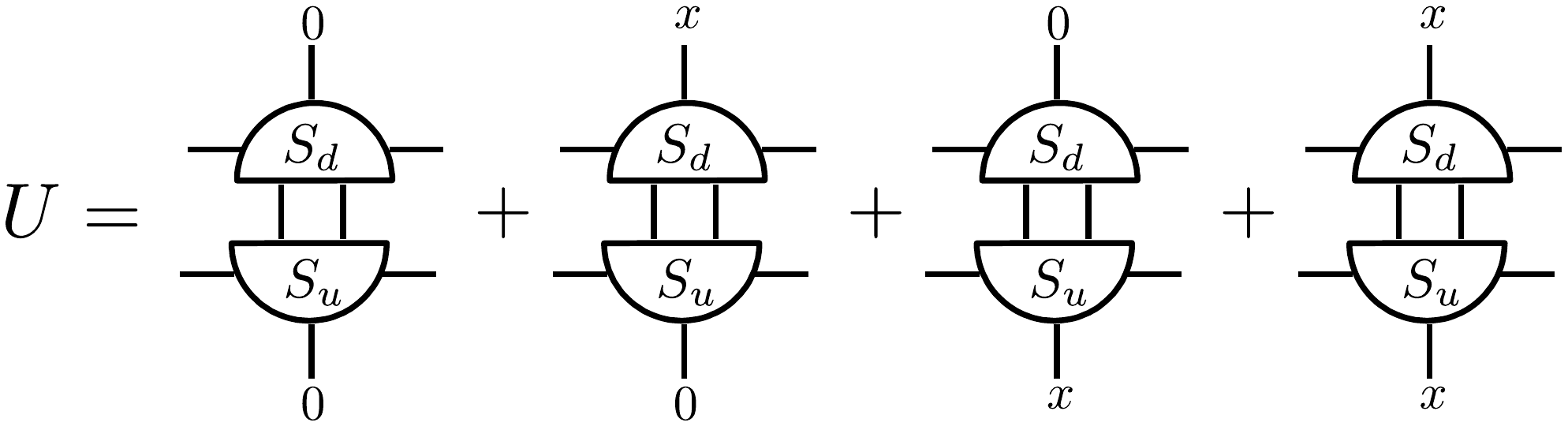}\,. 
		\label{Ufinal}
	\end{equation}
	In the second, third and fourth terms one or both contracted tensors are $O(\epsilon^{3/2})$, so those terms are $O(\epsilon^{3/2})$. For the first term
	we compute the $O(1)$ and $O(\epsilon)$ terms, using the definition \eqref{Su}. The $O(1)$ term is, after contracting with an isomorphism $J$ (step II.5), the high-temperature fixed point tensor. While $O(\epsilon)$ corrections could in principle arise as cross-terms between $O(1)$ piece of $S_u$ and $O(\eps)$ piece of $S_d$ and vice versa, luckily they vanish by the corner structure of the one-circle tensor.\footnote{The $O(\epsilon)$ part of the $S_u$ tensor is the diagram with a single one-circle tensor insertion in \eqref{Su}. Since the one-circle tensor only has corner components, we see that the top indices of this diagram are $0x$ (or $x0$ for the reflection). Hence this diagram cannot be contracted with the $O(1)$ part of the $S_d$ tensor, which has all indices 0.} 
    Nonzero corrections do appear in the next order $O(\eps^2)$. In particular the 0000 component becomes $\calN=1+O(\eps^2)$. Dividing by it, we get a tensor $A'$ satisfying A1 with $\eps$ replaced by $\eps^{3/2}$. \qed
	
	\subsection{Main result}
	
	Combining Propositions \ref{prop:type0}, \ref{prop:typeI}, \ref{prop:typeII} we get the following theorem which is our main result:
	
	\begin{theorem}\label{main}
		Let $A=A_* + \delta A$ be a normalized tensor ($A_{0000}=1$). It is possible to perform a tensor RG transformation followed by normalization which maps $A$ to a normalized tensor $A'=A_* + \delta A'$ with
		\beq
		\|\delta A'\|\le C \|\delta A\|^{3/2}\,.
		\eeq
	\end{theorem}
	
	The constant $C=O(1)$ could be extracted by following carefully the proofs of the propositions; we will not do it here.
	In particular for $\|\delta A\|< 1/C^2$ we have $\|\delta A'\|< \|\delta A\|$ and the iterates of the RG transformations will converge to $A_* $ super-exponentially fast.
	
	\section{Conclusions}\label{sec:concl}
	
	In this paper we rigorously formulated a
	tensor RG approach for the high-$T$ phase of 2D lattice models. Of course many other traditional techniques also work in the high-$T$ phase, such as the
	high-temperature expansion.
	Although at general temperatures the Wilson-Kadanoff
	RG has not even been defined rigorously, at high $T$ one can express it in a rigorous expansion, and show convergence to the high-temperature fixed point
	{\cite{Griffiths1979,Kashapov1980}}.
	Here we showed how a similar RG stability
	result can be obtained using tensor RG.
	
	In the future our work can be extended in several directions. In Appendix \ref{cluster} we discuss how standard cluster expansions techniques can be used to prove exponential decay of correlation functions when the tensor $A$ is close to the high temperature fixed point; we then discuss how the tensor RG might be combined with these expansions to expand the region in which exponential decay can be proven. Another direction for future work is to consider higher-dimensional lattices. 
	
	It would be even more interesting to extend tensor RG methods to other situations where traditional RG techniques become complicated. One such situation arises at low temperatures, where the traditional approach has been to apply RG in a contour representation, which introduces some complications {\cite{Gawedzki1987,Bricmont1988}}. In fact it is known that the Wilson-Kadanoff RG transformation in the spin representation becomes ill-defined at low temperatures (see {\cite{van_Enter_1993}} and {\cite{BRICMONT2001}} for a review).

	On the contrary, tensor RG is just as well defined at low as at high
	temperature. At low $T$ and with zero magnetic field, one expects convergence
	to the low-temperature fixed point described by the direct sum $A_{\ast}
	\oplus A_{\ast}$ of two high-temperature fixed point tensors. Such convergence
	is seen in the numerical studies (see Appendix \ref{sec:prior}). It should be
	possible, and interesting, to establish this rigorously using techniques of
	the present paper.
	
	Finally, let us discuss the critical temperature. Although Wilson-Kadanoff RG
	should be well defined at criticality (see {\cite{Tom}} for a recent
	discussion and numerical work), a robust algorithm to evaluate it precisely is
	lacking. On the contrary, tensor RG evaluation is in principle
	straightforward. The CDL tensor problem, discussed in Appendix \ref{CDL}, \
	requires one to complicate tensor RG transformations somewhat compared to the
	most basic versions, but they remain manageable.
	
	We are therefore optimistic that in the not too distant future the existence of a nontrivial tensor RG fixed point for the 2D Ising and other 2D lattice
	models will be proved.
	The proof will proceed in two steps. First, using an
	appropriate tensor RG transformation, one constructs numerically an
	approximate RG fixed point. Then, by functional analysis, one proves that an
	exact fixed point exists nearby. This would therefore be a computer-assisted
	proof, in the spirit of Lanford's construction of the Feigenbaum fixed point
	{\cite{lanford1982}}. We think this is a very interesting open problem.
	
	A similar proof for the critical point of the 3D Ising model would be even
	more exciting. This should wait until numerical tensor RG methods are able to
	approximate this fixed point with a good accuracy, which has not yet been
	achieved (see Appendix \ref{Gilt}).
	
	\section*{Acknowledgements}
	
	SR is supported by the Simons Foundation grants 488655 and 733758 (Simons Collaboration on the
	Nonperturbative Bootstrap), and by Mitsubishi Heavy Industries as an ENS-MHI
	Chair holder. 
	
	\section*{Data availability statement}

	Data sharing is not applicable to this article as no datasets were generated or analysed during the current study.

	\appendix
	
	\section{Tensor representation for the 2D Ising model}
	\label{sec:ising}

	In this appendix we show how to transform the partition function of the
	nearest-neighbor 2D Ising model on the square lattice with the Hamiltonian $H
	= -\beta \sum_{\langle i j \rangle} \sigma_i \sigma_j$, $\sigma_i = \pm 1$, to
	a tensor network built out of 4-tensors as in Fig.~\ref{fig-intro1}. We present two ways to perform this transformation.
	
	\subsection{Rotated lattice}
	
	This representation was used e.g.~in \cite{TEFR}. We start by rotating the spin lattice by 45 degrees. We now
	associate a tensor with every second square, tensor contractions taking place at the
	positions of Ising spins: 
	\beq
	\myinclude[scale=0.4]{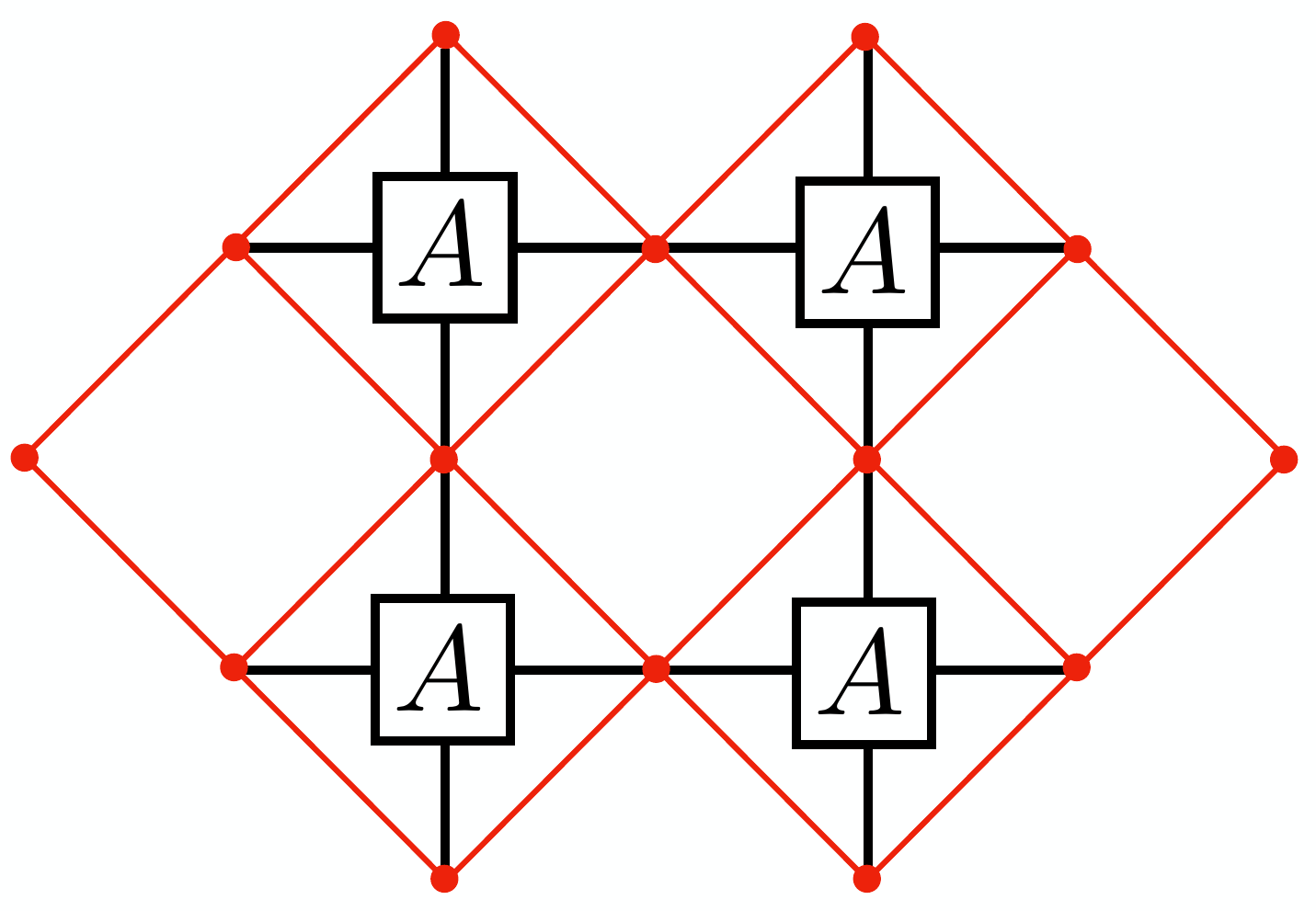}\ .
	\eeq
	The Hilbert space
	has two basis elements $| \sigma \rangle  $, $\sigma = \pm$. Assigning tensor components
	\beq
	\myinclude[scale=0.5]{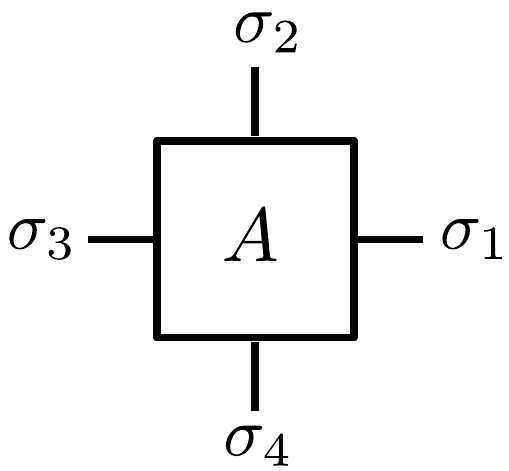} = e^{\beta (\sigma_1 \sigma_2
		+ \sigma_2 \sigma_3 + \sigma_3 \sigma_4 + \sigma_4 \sigma_1)}, 
	\eeq
	we reproduce the 2D Ising model partition function. In finite volume
	this gives the partition function with periodic boundary conditions on
	the lattice rotated by 45 degrees.
	
	Transforming the basis to the $\mathbb{Z}_2$ even ($0$) and odd ($1$) states
	\beq
	|  0 \rangle = \frac{1}{\sqrt{2}} (| +  \rangle 
	+ | -  \rangle  ), \qquad |  1
	\rangle = \frac{1}{\sqrt{2}} (| +  \rangle   - | - 
	\rangle  ), 
	\eeq
	the nonzero tensor components become:
	\begin{gather} 
		A_{0000} = \cosh (4 \beta) + 3,\quad
		A_{0101}=A_{1111} = \cosh (4 \beta) - 1,\quad A_{0011} = \sinh (4 \beta),
		\label{rep1}
	\end{gather}
	and rotations thereof. By $\mathbb{Z}_2$ invariance, they all
	have an even number of $1$ indices.
	
	\subsection{Unrotated lattice}
	
	Alternatively, one can perform the transformation without rotating the lattice \cite{HOTRG,GILT}. In finite volume, this method gives the partition function with the usual periodic boundary conditions. We start by representing $W=e^{\beta \sigma_1 \sigma_2}$, viewed as a two-tensor in the Hilbert space with basis elements $| \sigma \rangle  $, $\sigma = \pm$, as a contraction of two tensors $W=M M^T$ with
	\beq
	M=\begin{pmatrix} \sqrt{c} &  \sqrt{s}\\
		\sqrt{c} &  -\sqrt{s} \end{pmatrix}\,,
	\eeq
	where $c=\cosh \beta$, $s=\sinh\beta$. Graphically
	\beq
	\myinclude[scale=0.25]{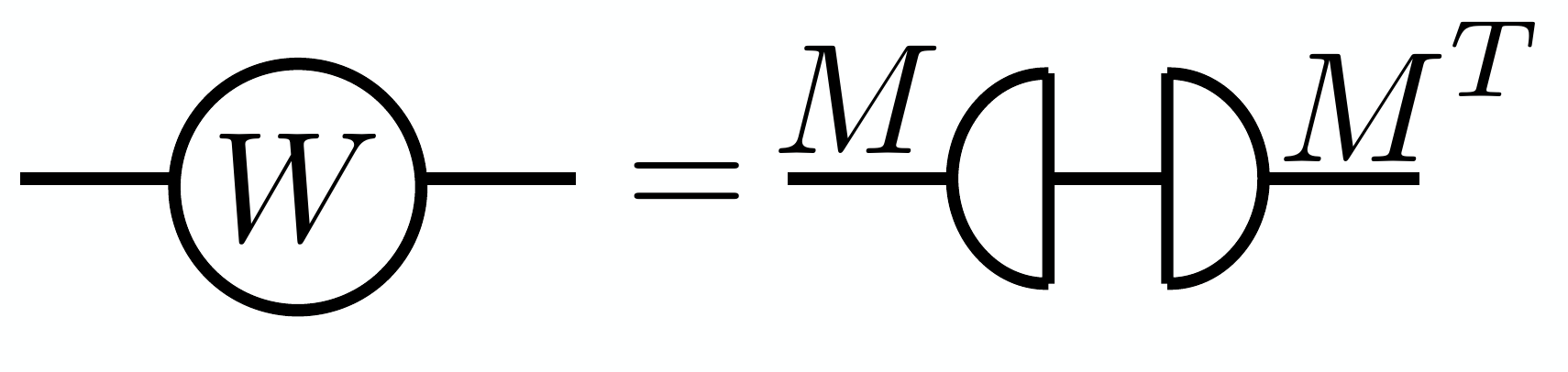}\ .
	\eeq
	Using 0 and 1 to index the columns of the $M$ matrix, equations
	\beq
	M_{+0}=M_{-0},\quad M_{+1}=-M_{-1}
	\eeq
	mean that the states 0,1 are, respectively, $\mathbb{Z}_2$ even and odd. 
	A tensor network representation of the partition function is obtained by defining
	\beq
	A_{ijkl}=M_{+i} M_{+j}M_{+k} M_{+l} + M_{-i} M_{-j}M_{-k} M_{-l}
	\eeq
	or graphically
	\beq
	\myinclude[scale=0.25]{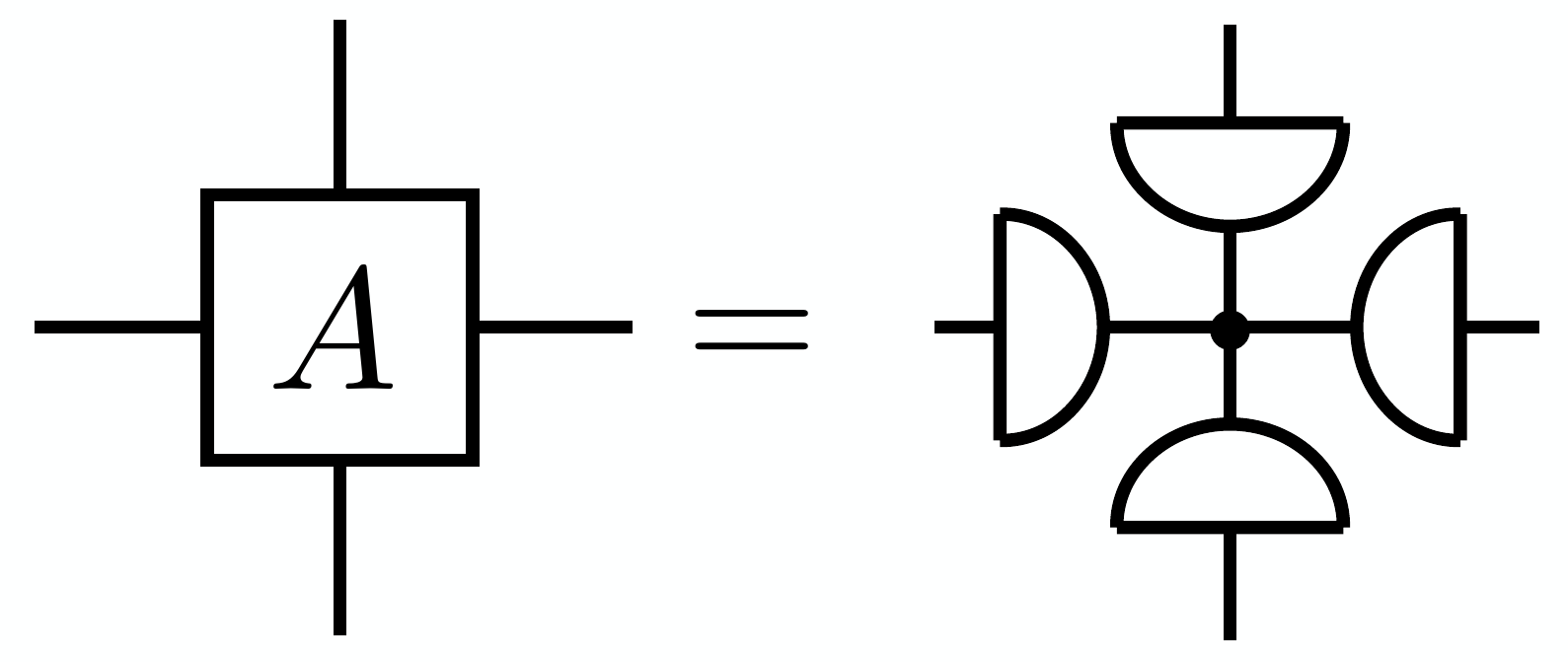}\ .
	\eeq
	The nonzero components are
	\beq
	A_{0000}  = 2 c^2, \quad A_{0011} = A_{0101}=2cs, \quad A_{1111}=2s^2,
	\label{rep2}
	\eeq
	and rotations thereof.
	
	\subsection{Stability of the high-temperature fixed point}
	
	For both representations \eqref{rep1} and \eqref{rep2}, tensor $A$ in the limit $\beta \rightarrow 0$ becomes proportional to the high-temperature fixed point tensor $A_{\ast}$, as component $A_{0000}$ tends to a constant and all others to zero. Theorem \ref{main} allows us to conclude that for sufficiently small temperatures $\beta \leqslant \beta_0$ the tensor RG iterates of $A$ will converge to $A_{\ast}$.

	\section{Exponential decay of correlators}
	\label{cluster}

{
  The results in this paper do not prove exponential decay of correlation functions when the tensor $A$ is close to the high temperature fixed point. This is not a shortcoming of the tensor network RG approach compared to the Wilson-Kadanoff RG. For the Wilson-Kadanoff RG one can prove that the RG map is rigorously defined and its iterates converge to the trivial fixed point if the starting Hamiltonian is well inside the high temperature phase {\cite{Griffiths1979,Kashapov1980}}. However, this by itself does not imply exponential decay of correlations. Of course, if the Hamiltonian is well inside the high temperature phase then standard high temperature expansion methods can be used to prove exponential decay of correlations. Similarly, standard polymer expansion methods can be used to prove exponential decay of correlations for tensor networks when the tensor $A$ is sufficiently close to the high temperature fixed point. We will give a brief sketch of how this is done. This appendix is intended for readers familiar with such techniques, so we will be brief.

        We assume that $A=A_*+\delta A$ where $A_*$ denotes the high temperature fixed point as before and $||\delta A||$ is small. We use this equation to expand the partition function (the contraction of the tensor network) into a sum of terms where at each vertex in the lattice there is either $A_*$ or $\delta A$.  We call a polymer any connected subset $\gamma$ of lattice sites. The set of lattice sites with $\delta A$ can then be written as the disjoint union of a collection of polymers. The value of this term in the expansion of the partition function is equal to the product of the activities of these polymers where the activity $w(\gamma)$ of a polymer $\gamma$ is the value of the contraction of the tensor network when there is a $\delta A$ at each site in $\gamma$ and an $A_*$ at the sites not in $\gamma$. We then have the bound $|w(\gamma)| \le \|\delta A\|^{|\gamma|}$ where $|\gamma|$ is the number of vertices in $\gamma$. Then standard polymer expansion methods can be applied if $\|\delta A\|\le \eps_{\rm 0}$ \cite{kotecky_preiss_1986}. Here $\eps_{\rm 0}$ is some small number which will come out from the standard criteria of the polymer expansion convergence.


Where does tensor RG enter this picture? Our Theorem \ref{main} proves that if $\|\delta A\|\le \eps_{\rm TRG} =1/C^2$ then we can apply tensor RG repeatedly to make $\|\delta A\|$ arbitrarily small. We have not worked out in this paper what $\eps_{\rm TRG}$ or $\eps_0$ are.  If $\eps_{\rm TRG}>\eps_0$, then the polymer expansion techniques described above cannot be used directly to show that all these tensor networks are in the high-temperature phase. But combining the two we could do this as follows. 

In this paper we discussed tensor RG for the partition function. Correlation function can be represented in terms of a tensor network in which some of the $A$ tensors are replaced by other tensors. (The contraction of this network must be divided by the partition function to get the correlation function.) We will refer to these other tensors as $B$ tensors. When one applies tensor RG to the network with some $B$ tensors, these $B$ tensors will be changed as well as the $A$ tensors. We do not have any control over how the norm of the $B$ tensors may change, but that does not matter. After a finite number of iterations of the RG map the tensor $A$ will be sufficienly close to the high temperature fixed point that we can apply a polymer expansion to get exponential decay of the correlation functions. 

A calculation of $\eps_{\rm TRG}$ based directly on the proof in this paper may not give a value greater than $\eps_0$. Nonetheless we hope that the value of
$\eps_{\rm TRG}$ can be improved (i.e., increased) considerably by taking advantage of the fact that the tensor RG map happens locally in space. So there is the possibility that the tensor RG can be used to expand the range of temperatures for which we know that the Ising model (and perturbations of it) are in the high temperature phase. 
}

	\section[Prior literature on tensor RG]{Prior literature on tensor RG\footnote{In this paper the expressions ``tensor RG'' or ``tensor network RG'' denote the whole circle of ideas of performing RG using tensor networks, not limited to the TRG and TNR algorithms described in this section.}}

	\label{sec:prior}
	
	\subsection{TRG and its variants}\label{TRG}
	
	Tensor network approach to RG of statistical physics systems was first
	discussed in 2006 by Levin and Nave {\cite{Levin:2006jai}}. The precise
	method they introduced, called Tensor RG (TRG), consists in representing
	tensor $A$ as the product $S S^T$, graphically
	\begin{equation}
		\myinclude[scale=0.25]{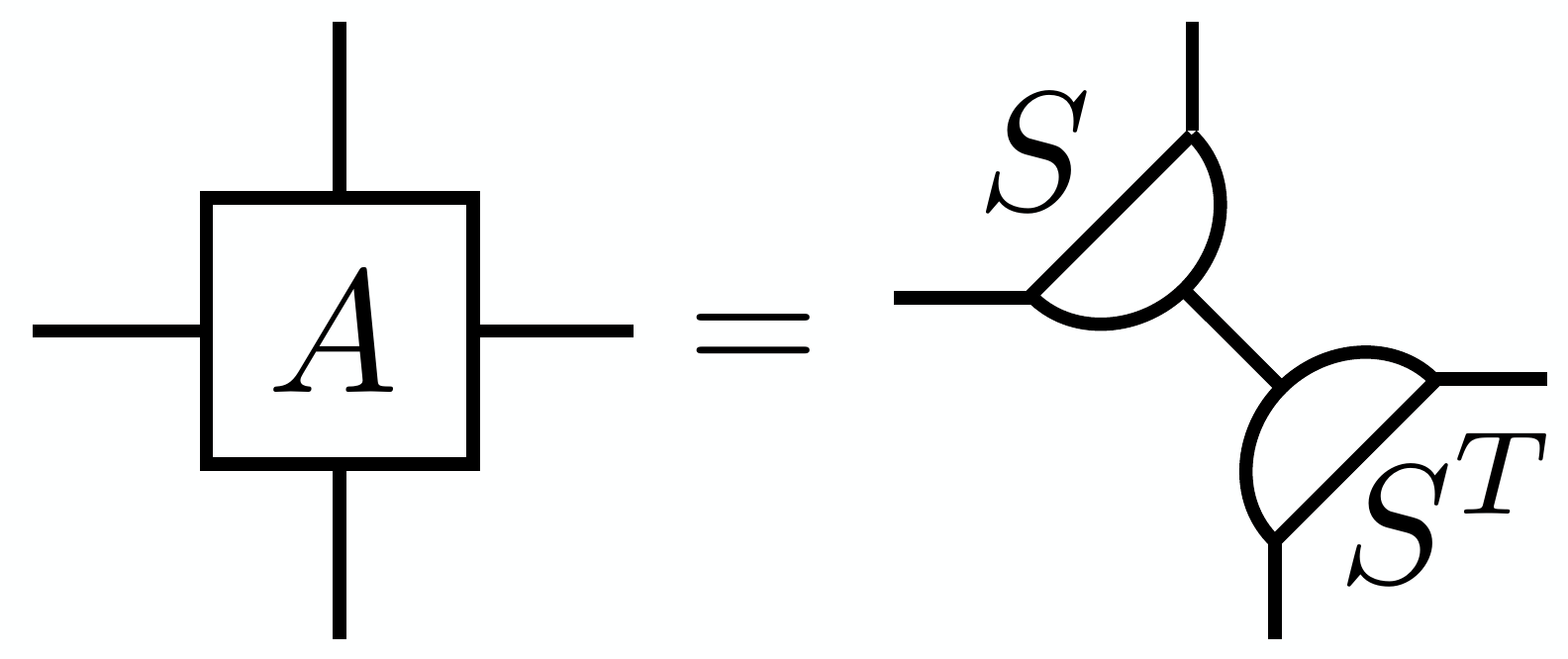}\label{AS}
	\end{equation}
	and then defining a new tensor by contracting four $S$ tensors:
	\begin{equation}
		\myinclude[scale=0.25]{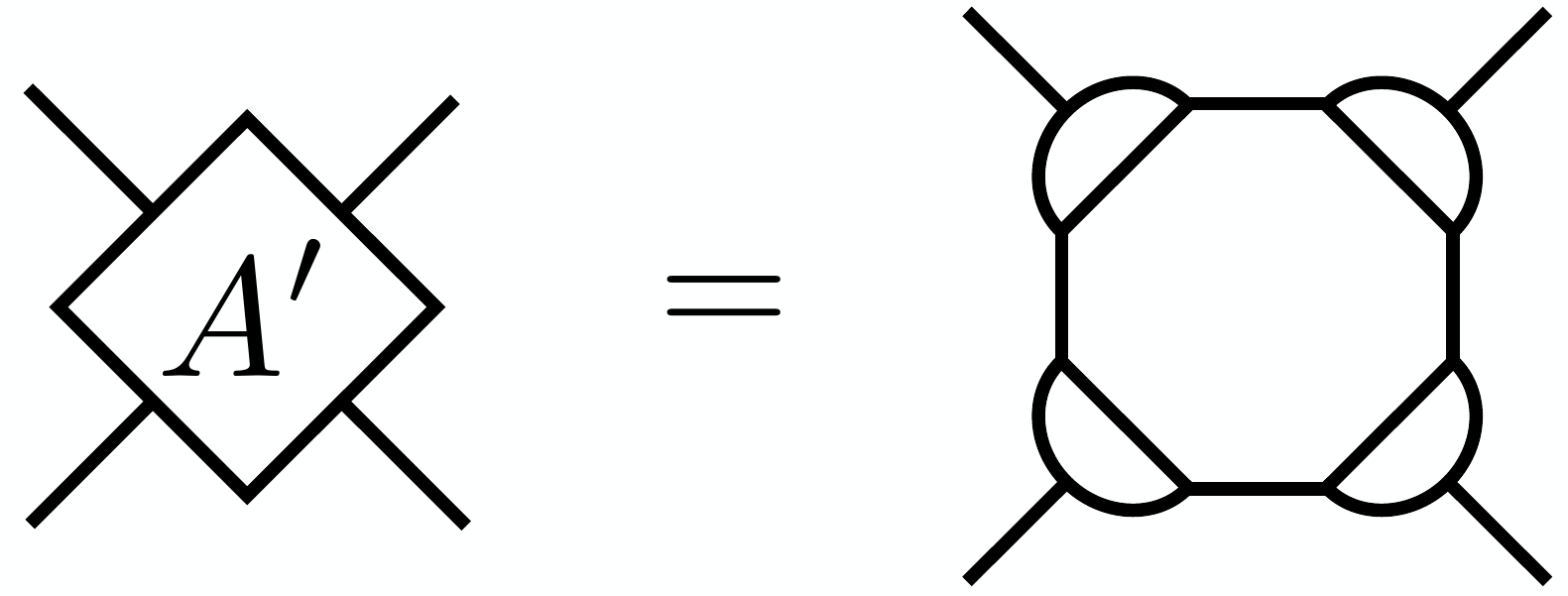}
	\end{equation}
	Decomposition {\eqref{AS}} is obtained from the singular value decomposition
	(SVD) $A = U \Lambda U^T$ with $\Lambda$ diagonal and $S = U \sqrt{\Lambda}$.
	To keep the bond dimension $D$ finite, the singular value decomposition is
	truncated to the largest $D$ singular values out of the total of $D^2$. Ref.~{\cite{Levin:2006jai}
		also defined a similar procedure for the honeycomb lattice. TRG can be used to evaluate numerically the free energy
		of the 2D Ising model as a function of the temperature and of the external
		magnetic field on a large finite lattice (reducing it to a one-site lattice by
		a finite number of TRG steps), and then extract the infinite volume limit by
		extrapolation. Using bond dimensions up to $D = 34$, Ref.~{\cite{Levin:2006jai} attained excellent
			agreement with the exact result except in a small neighborhood of the critical
			point.
			
			In the introduction we mentioned a simple tensor network RG based on Eq. {\eqref{A1}}, which was also the basis for type I RG step in section \ref{sec:typeI}. This procedure was first carried out
			in {\cite{HOTRG}},\footnote{For illustrative purposes, Eq. {\eqref{A1}} also
				appeared in {\cite{TEFR}}.} called there HOTRG since it used
			higher-order SVD in the bond dimension truncation step. More precisely, HOTRG first contracts tensors
			pairwise in the horizontal direction, and then pairwise in the vertical
			direction. HOTRG agrees with the exact 2D Ising free energy even better than
			TRG. Applied to the 3D Ising, it gives results in good agreement with Monte
			Carlo. The accuracy of TRG and HOTRG is further improved by taking into
			account the effect of the ``environment'' to determine the optimal truncation,
			in methods called SRG {\cite{SRG,SRG1}} and HOSRG {\cite{HOTRG}}.
			
			\subsection{CDL tensor problem}\label{CDL}
			
			Although TRG and its variants performed very well in numerical free energy
			computations, it was noticed soon after its inception  {\cite{Levin-talk}} that the method has a
			problem. There is a class of tensors describing lattice
			models with degrees of freedom with ultra-short-range correlators only, which
			are not coarse-grained away by the TRG method and its above-mentioned variants
			but passed on to the next RG step unchanged, being exact fixed points of the
			TRG transformation. These ``corner double line'' (CDL) tensors (named so in
			{\cite{TEFR}}) have a bond space with two indices and the components
			\begin{equation}
				A_{i_1 i_2, j_1 j_2, k_1 k_2, l_1 l_2} = M^{(1)}_{i_1 j_2} M^{(2)}_{j_1 k_1}
				M^{(3)}_{k_2 l_1} M^{(4)}_{l_2 i_2}
			\end{equation}
			or graphically
			\begin{equation}
				\myinclude[scale=0.25]{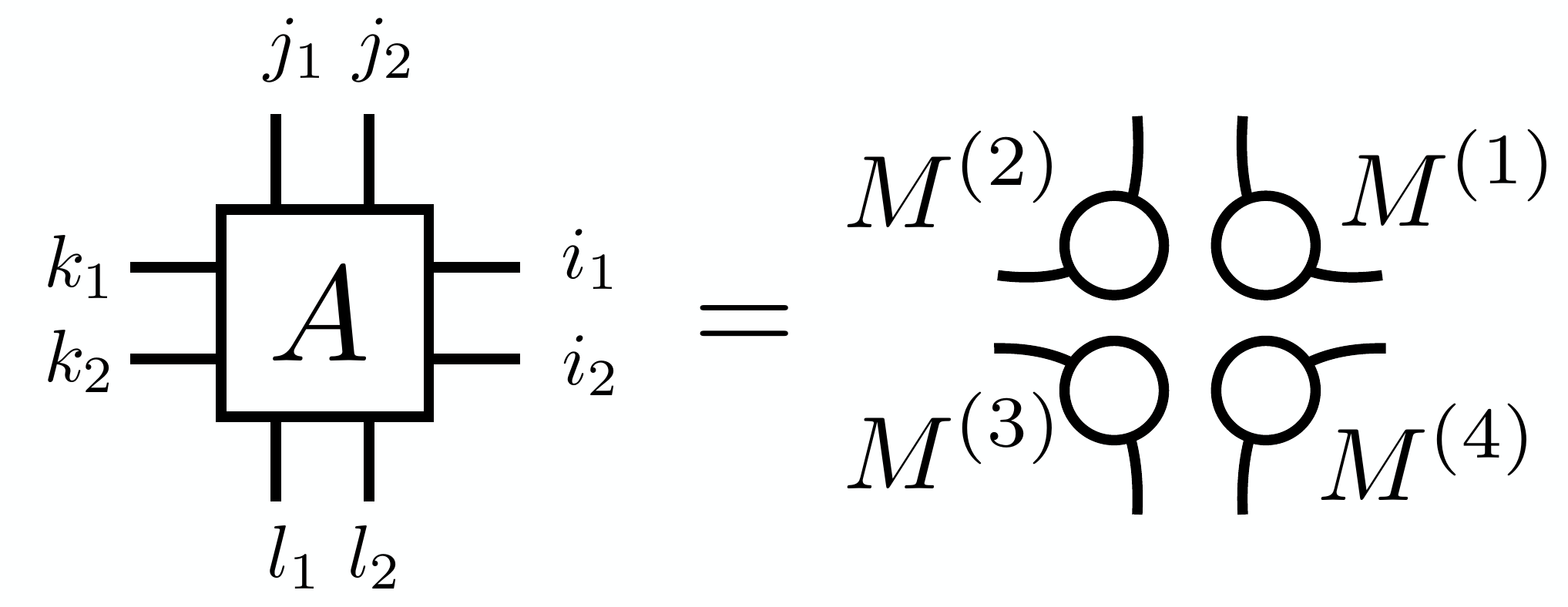}
			\end{equation}
			with arbitrary matrices $M^{(1)}, \ldots, M^{(4)}$. CDL tensors present a
			problem both conceptually and practically:
			\begin{itemize}
				\item Conceptually, as we would like to describe every phase by a single fixed point, not by a class of equivalent fixed points
				differing by a CDL tensor.
				
				\item Practically, since every numerical tensor RG computation is performed
				in a space of finite bond dimension, limited by computational resources. It
				is a pity to waste a part of this space on uninteresting
				short-range correlations described by the CDL tensors. This becomes
				especially acute close to the critical point. 
			\end{itemize}
			We proceed to discuss algorithms trying to address this problem.
			
			\subsection{TEFR}\label{TEFR}
			
			TEFR (tensor entanglement filtering renormalization) was the first algorithm
			aiming to resolve the problem of CDL tensors {\cite{TEFR}}. It consists of the
			following steps:
			\begin{enumerate}
				\item Perform SVD to represent $A$ as in Eq. {\eqref{AS}}, truncating to the
				largest $D$ singular values so that the representation is approximate, where
				$D$ is the original bond dimension. Tensor $S$ thus has dimension $D \times D
				\times D$.\footnote{Two three-tensors appearing in the decomposition
					{\eqref{AS}} may in general be different but we will treat them here as identical
					for simplicity. The same comment applies to tensors
					$S_1$ and $S_2$ below.}
				
				\item Find a tensor $S_1$ of dimension $D \times D_1 \times D_1$ with $D_1 <
				D$ so that\footnote{More generally, the four tensors marked by $S_1$ don't have to be identical, but we will assume that they are identical to simplify the notation.}
				\begin{equation}
					\myinclude[scale=0.3]{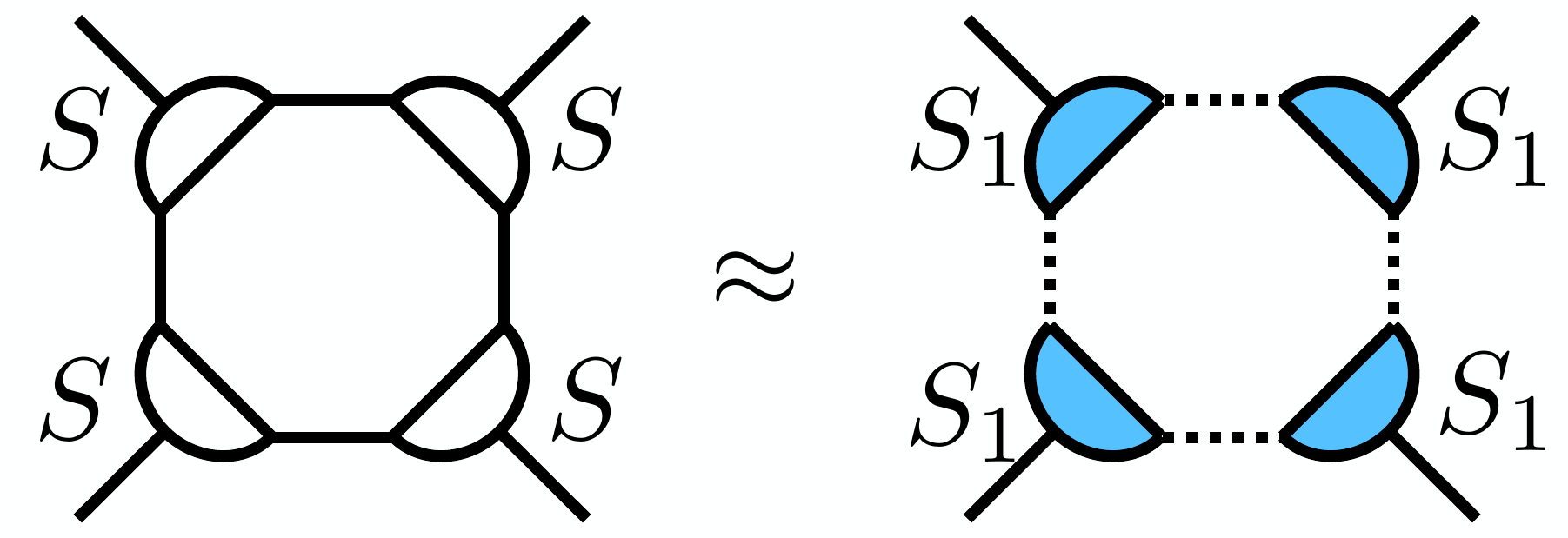}\ , \label{TEFR2}
				\end{equation}
				where dashed bonds have dimension $D_1$.
				
				\item Recombine tensors $S_1$ to form a four-tensor $A_1$ with
				dimension $D_1 \times D_1 \times D_1 \times D_1$:
				\begin{equation}
					\myinclude[scale=0.3]{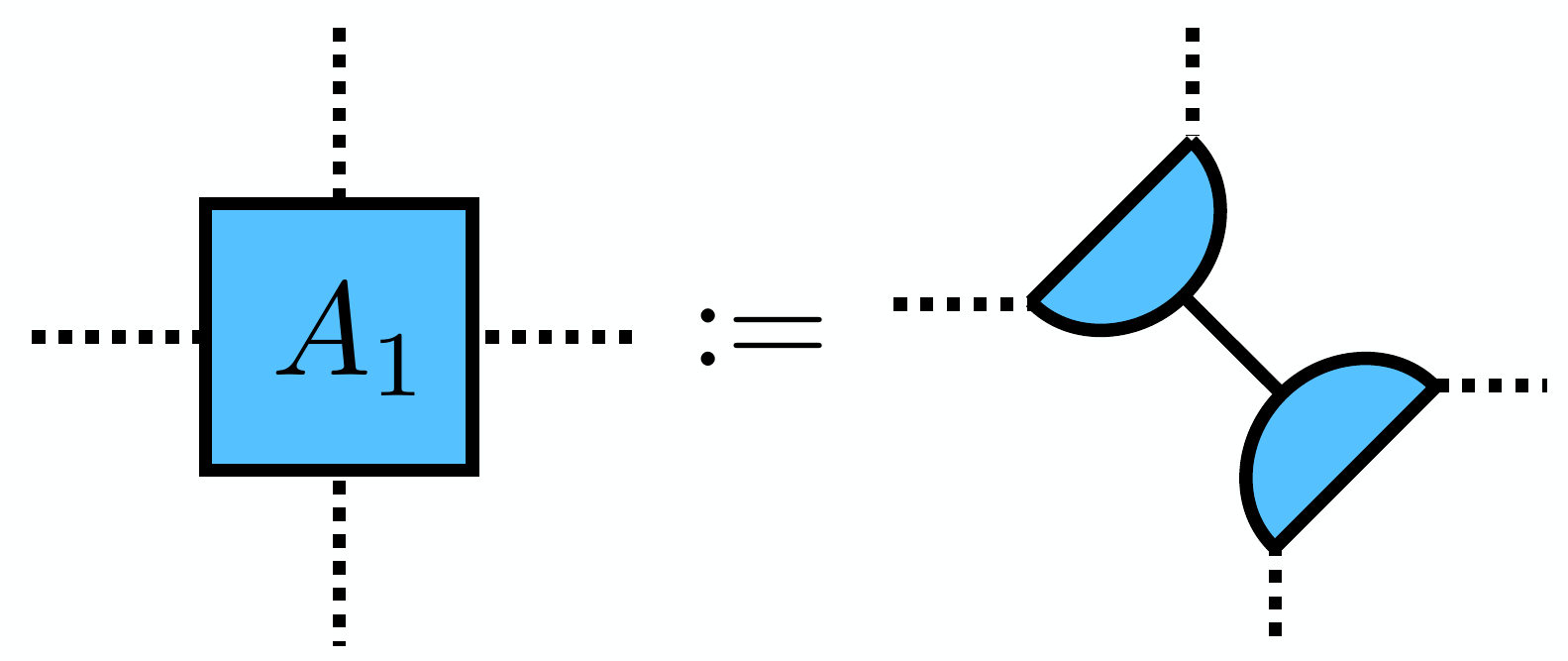}\ .
				\end{equation}
				\item Perform an SVD to represent $A_1$ along the "other diagonal":
				\begin{equation}
					\myinclude[scale=0.3]{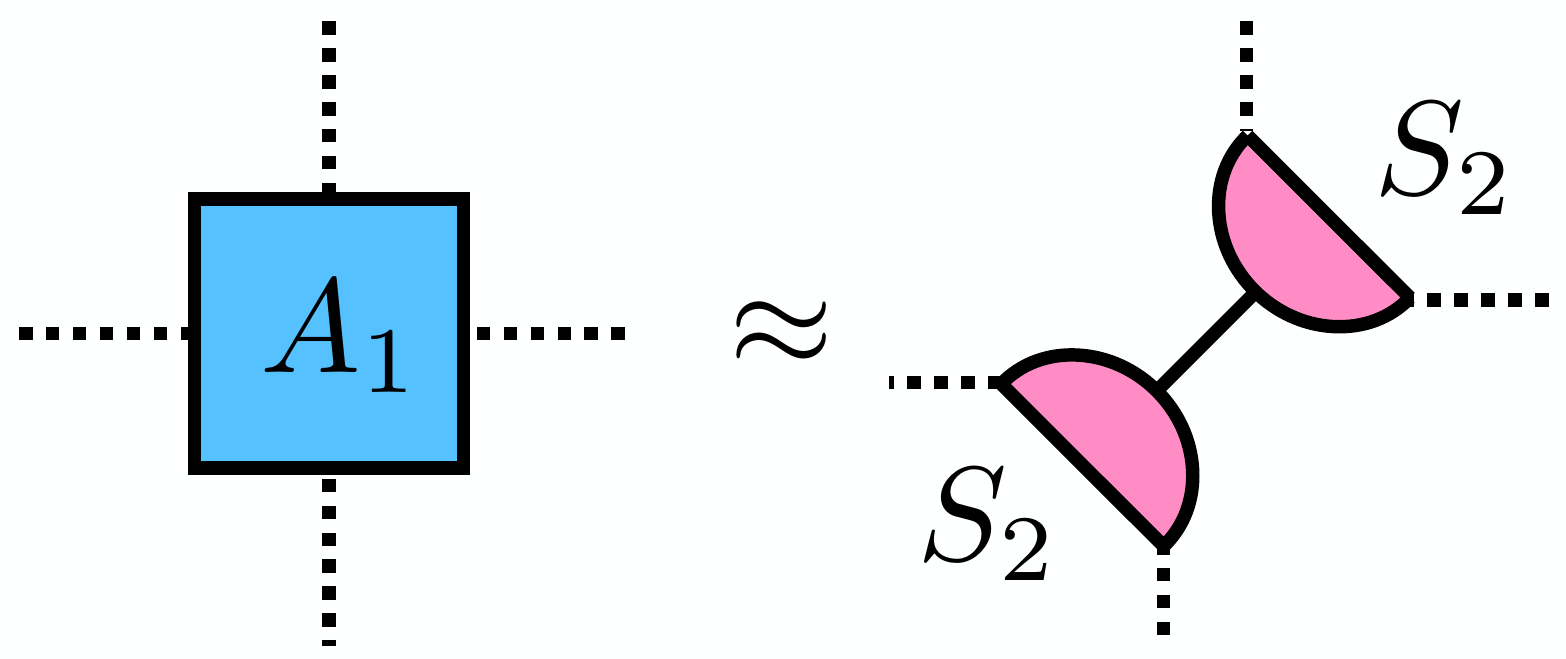}\ .
				\end{equation}
				\item Contract $S_2$ tensors to form tensor $A'_{}$, the final result of the
				RG step:
				\begin{equation}
					\myinclude[scale=0.3]{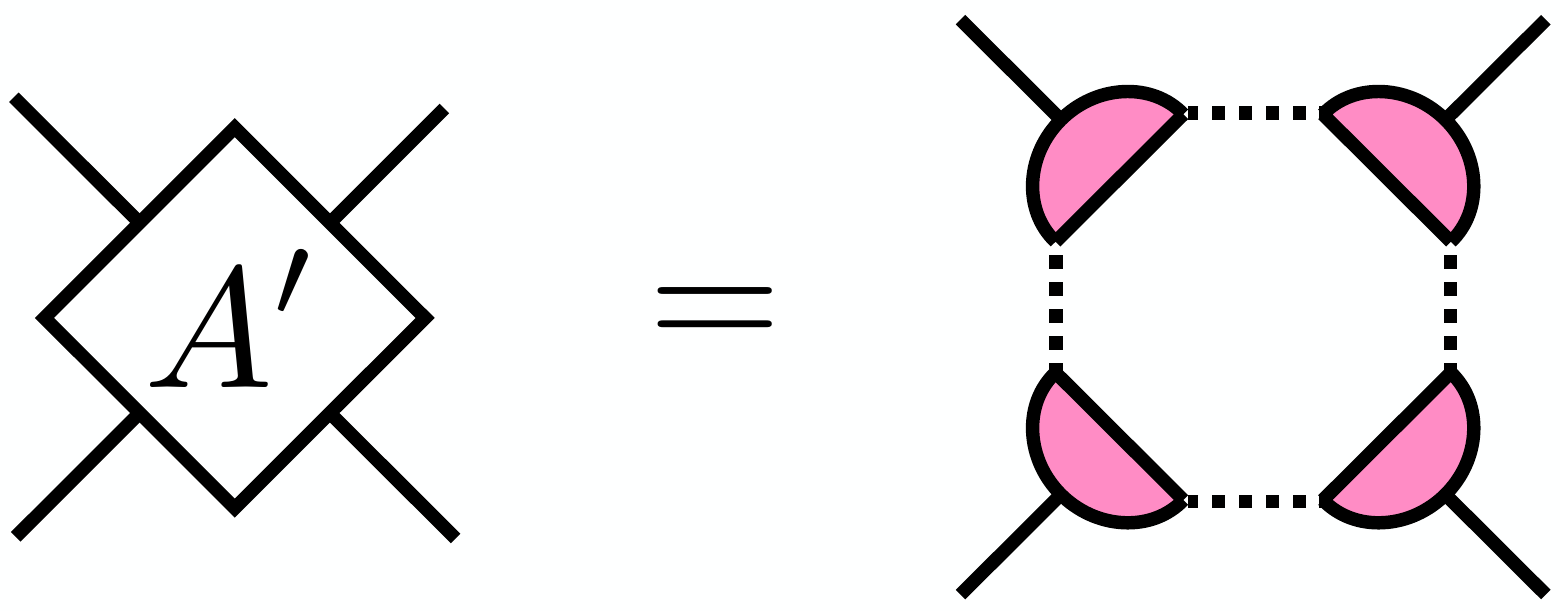}\ .
				\end{equation}
			\end{enumerate}
			The rationale behind this algorithm is that it reduces an arbitrarily large CDL tensor to a tensor of dimension one in a single RG step {\cite{TEFR}}. It is less clear what happens for 
			tensors which are not exactly CDL. Ref.
			{\cite{TEFR}} gives several procedures for how to perform the key Step 2 for such tensors, but it's not
			clear if these are fully successful. Numerical results of {\cite{TEFR}} for
			the 2D Ising model are inconclusive. TEFR RG flow converges to
			the high-temperature fixed point tensor $A_{\ast}$ for $T > 1.05 T_c$, to the
			low-temperature fixed point tensor $A_{\ast} \oplus A_{\ast}$ for $T < 0.98
			T_c$, while in the intermediate range around $T_c$ it converged to a
			$T$-dependent tensor with many nonzero components.
			
			Gilt-TNR algorithm (App.~\ref{Gilt} below) improves on TEFR by providing a
			more robust procedure for Step 2.
			
			\subsection{TNR}\label{TNRG}
			
			TNR (tensor network renormalization) algorithm by Evenbly and Vidal
			{\cite{Evenbly-Vidal}} was perhaps the first method able to address
			efficiently the CDL tensor problem. It gives excellent results when applied to
			the 2D Ising model {\cite{Evenbly-Vidal}}, including the critical point. It runs as follows:\footnote{We
				present the algorithm rotated by 90 degrees compared to
				{\cite{Evenbly-Vidal}}.}
			\begin{enumerate}
				\item Choose a unitary transformation of $V \otimes V$ (``disentangler'').
				Inserting $u u^{\dagger}$ into the tensor network does not change its value:
				\begin{equation}
					\myinclude[scale=0.4]{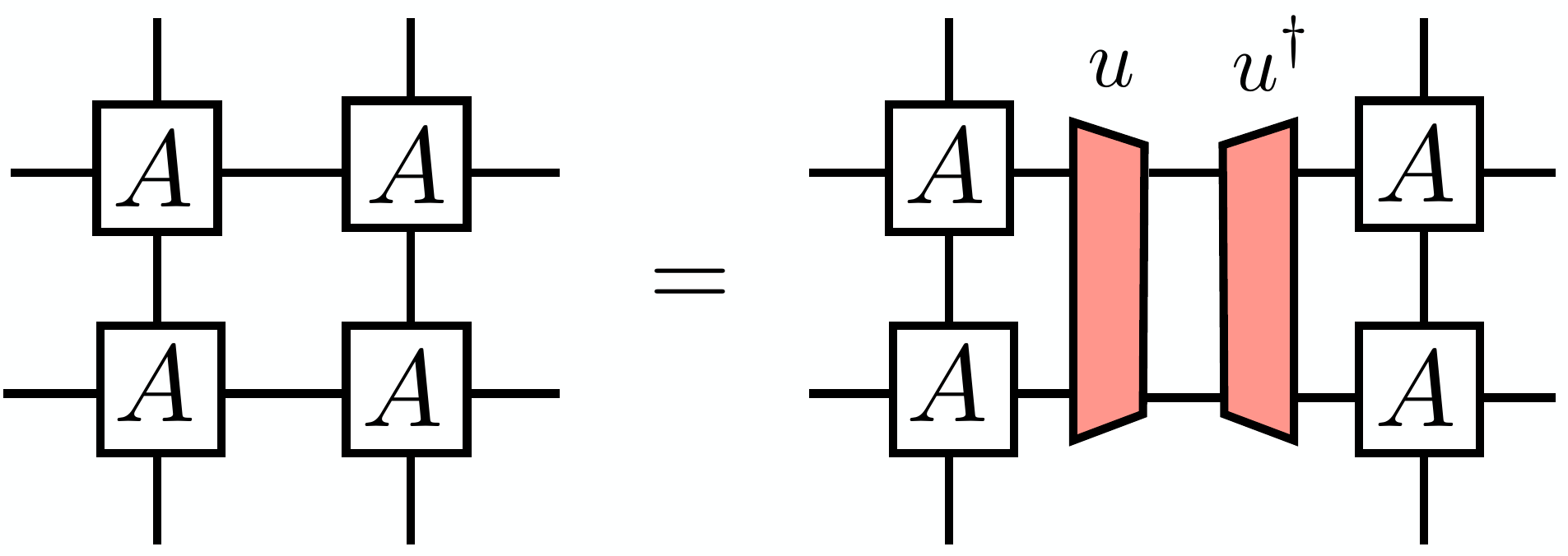}\ .
				\end{equation}
				\item Choose a couple of isometries $v_1,v_2 : V \rightarrow V \otimes V$. Optimize $u$, $v_1$,
				$v_2$ so that
				\begin{equation}
					\myinclude[scale=0.4]{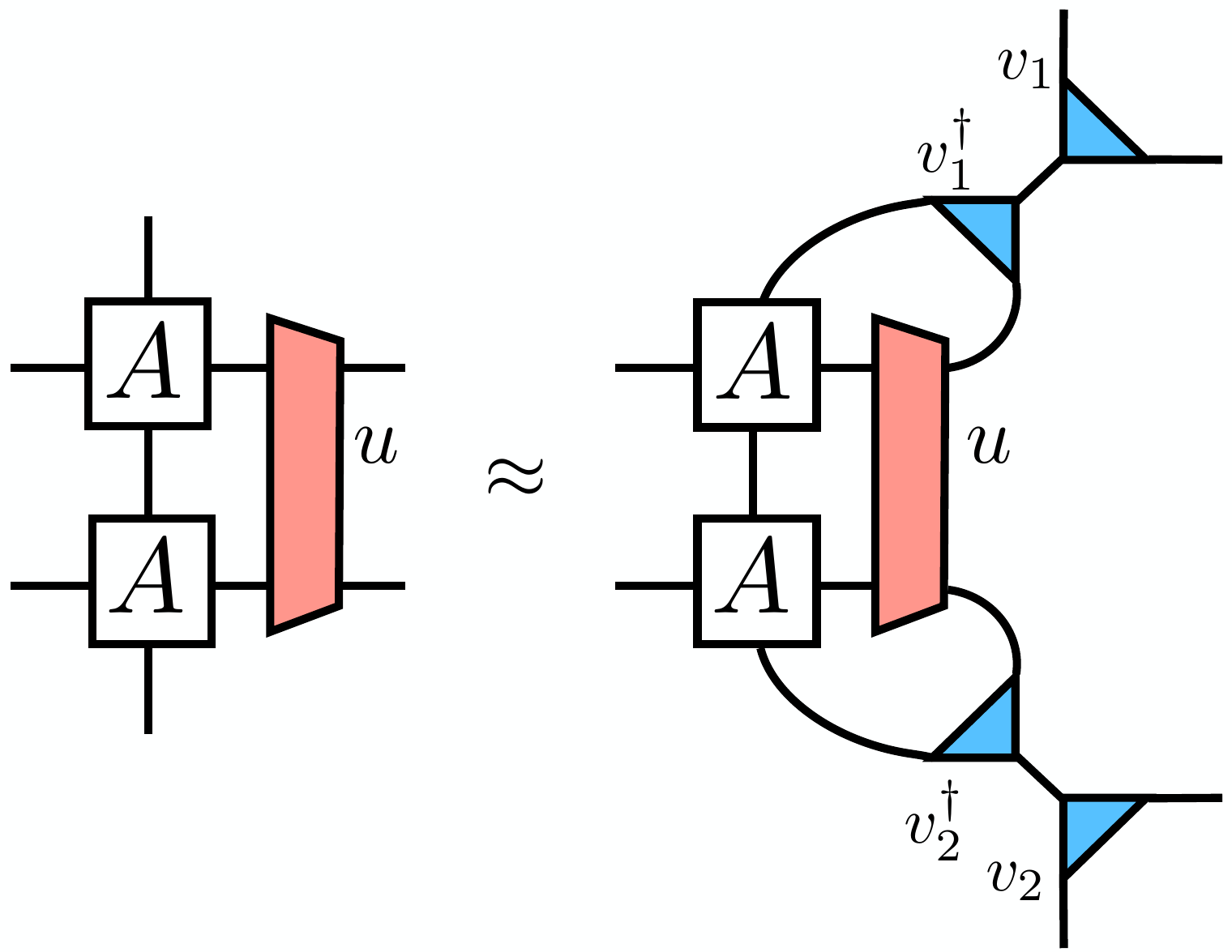}\ . \label{uv}
				\end{equation}
			      \item Rearrange the tensor network as
				\begin{equation}
					\myinclude[scale=0.6]{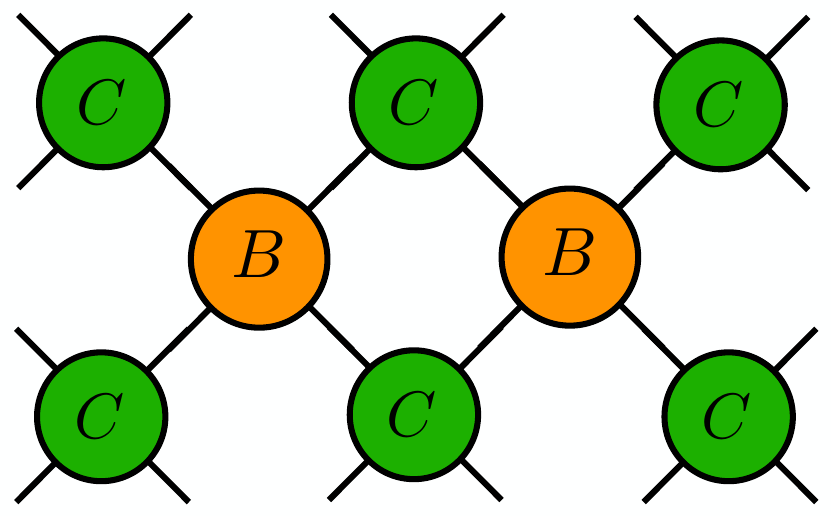}\ ,
				\end{equation}
				where (each blue triangle stands for one of the $v_i$, $v^\dagger_i$ tensors)
				\begin{equation}
					\myinclude[scale=0.6]{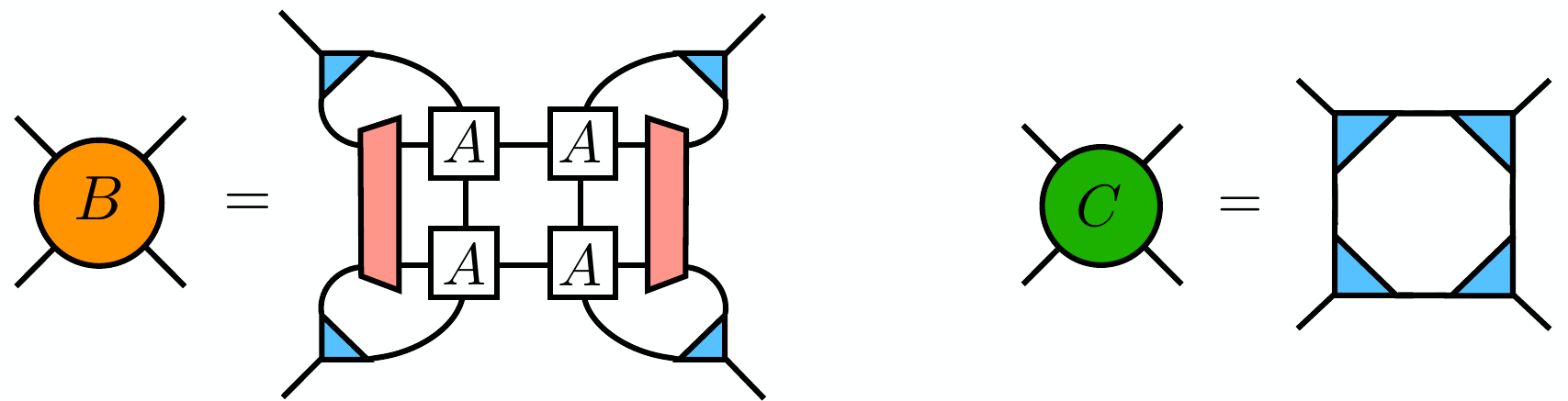}\,.
				\end{equation}
				\item Perform an (approximate because truncated) SVD for the $B$ and $C$
				tensors:
				\begin{equation}
					\myinclude[scale=0.6]{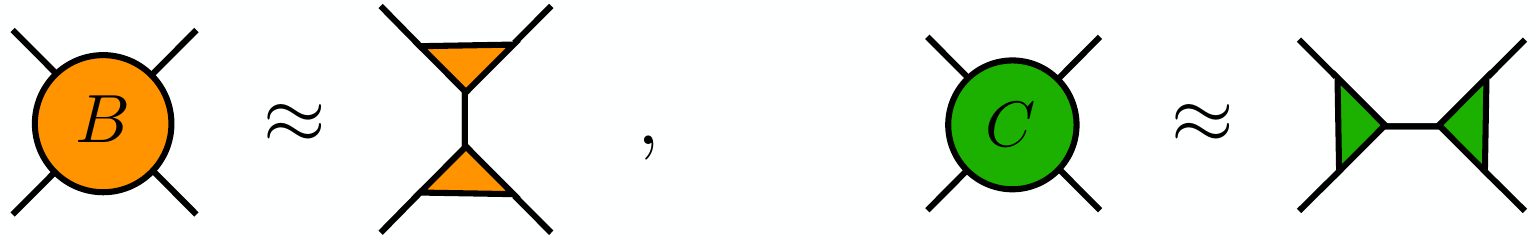}\,. \label{BC}
				\end{equation}
				\item Finally define the end result of the RG step as
				\begin{equation}
					\myinclude[scale=0.6]{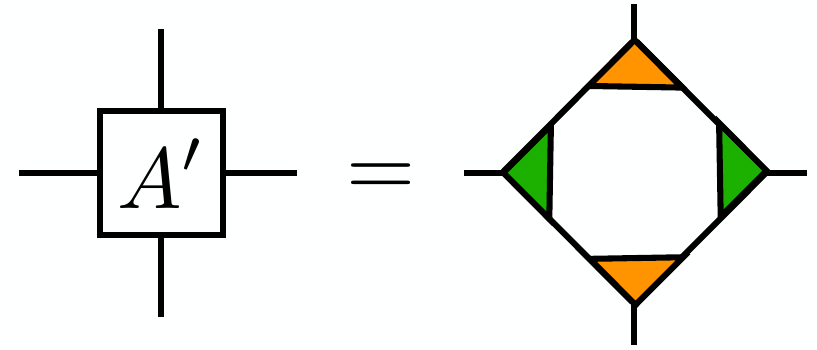}\,. \label{A1TNR}
				\end{equation}
			\end{enumerate}
			Our type II RG step (section \ref{sec:typeII}) was inspired by the TNR
			algorithm. The main differences and similarities are as
			follows:
			\begin{itemize}
				\item We work in an infinite-dimensional Hilbert space.
				
				\item Our disentanglers $R$ don't have to be unitary. We insert $R R^{-
					1}$ instead of $u u^{\dagger}$.
				
				\item We do not choose $u$ and $v_1,v_2$ to minimize the error in {\eqref{uv}}. Rather,
				our $v_1$, $v_2$ are the identity on $V \otimes V$, while $u$ is chosen to
				``disentangle'' certain correlations between the upper and lower part of TNR
				$B$ (whose section \ref{sec:typeII} analogue is called $S$), by killing the
				$O (\epsilon^2)$ part of the ``dangerous diagram'' in {\eqref{danger}}.
				Since our $v$'s are the identity, our $C$ is simply a product of Kronecker
				deltas, realizing a contraction among the indices of $B$ tensors
				
				\item Representation $S = S_u S_d$ in {\eqref{II3}} plays a role similar
				to the representation of $B$ in {\eqref{BC}} except that our representation
				does not involve any truncation and is not required to be an SVD.
				
				\item Our Eqs. {\eqref{Udef}} and {\eqref{II5}} are analogous to the
				TNR definition {\eqref{A1TNR}}. 
			\end{itemize}
			Further algorithms aiming to resolve the CDL problem include Loop-TNR
			{\cite{LoopTNR}} and TNR$_+$ {\cite{Bal:2017mht}}.
			
			\subsection{Gilt-TNR}\label{Gilt}
			
			Gilt(graph independent local truncation)-TNR {\cite{GILT}} follows the same
			steps as TEFR from Appendix \ref{TEFR}. It improves on TEFR by providing a very
			efficient way to perform Step 2, reducing dimensions of the contraction bonds
			in the l.h.s.~of {\eqref{TEFR2}} one bond at a time. See {\cite{GILT}} for
			details.
			
			Gilt-TNR is perhaps the simplest tensor network algorithm able to deal with
			the CDL problem. In addition to the usual 2D Ising benchmarking, Ref.~{\cite{GILT}} showed some preliminary results concerning its performance in 3D.
			It would be interesting to see a computation of the 3D Ising critical exponents by
			this or any other tensor RG method.
			
			See also {\cite{Lyu:2021qlw}} for a method combining Gilt with HOTRG from Appendix \ref{TRG}.

			\small
			\bibliographystyle{utphys}
			\bibliography{TN}

		\end{document}